\documentclass[journal]{IEEEtran}

\usepackage[utf8]{inputenc}
\usepackage[T1]{fontenc}
\usepackage{microtype}

\usepackage{amsmath, amsfonts, amssymb, mathtools}

\usepackage[most]{tcolorbox}
\usepackage{booktabs, multirow, tabularx, makecell, colortbl, threeparttable}
\usepackage{pifont}
\newcolumntype{L}{>{\arraybackslash}X}

\renewcommand*{\arraystretch}{1.5}
\newcommand{\cmark}{{\color{green!60!black}\ding{51}}}  
\newcommand{\xmark}{{\color{red!70!black}\ding{55}}}     
\newcommand{\wmark}{{\color{orange!85!black}\ding{115}}} 

\definecolor{tabred}{RGB}{230,36,0}
\definecolor{tabgreen}{RGB}{0,116,21}
\definecolor{taborange}{RGB}{250,124,30}
\definecolor{tabyellow}{RGB}{251,253,169}
\definecolor{babypink}{rgb}{0.96, 0.76, 0.76}
\definecolor{bananamania}{rgb}{0.98, 0.91, 0.71}
\definecolor{cambridgeblue}{rgb}{0.64, 0.76, 0.68}
\definecolor{lightgray}{gray}{0.96}

\usepackage{graphicx}
\usepackage{float}
\usepackage{caption}
\usepackage{subfigure}
\usepackage[figuresright]{rotating}
\usepackage[graphicx]{realboxes}

\usepackage[ruled,vlined]{algorithm2e}
\usepackage{algpseudocode}

\usepackage{xcolor}
\usepackage{enumitem}
\usepackage{siunitx}
\usepackage{indentfirst}

\usepackage{pifont}

\newcommand{\btc}{\textcolor{orange}{\rotatebox{345}{\textbf{B}}}\kern-0.4em}
\newcommand{\brc}{\textcolor{magenta}{\mathsf{B}\kern-0.3em\raisebox{.2ex}{|}}}

\usepackage{tikz}
\usetikzlibrary{arrows.meta, automata, positioning, calc, shapes, fit}
\usepackage{fontawesome5}

\usepackage{listings}
\lstdefinestyle{customjson}{
    backgroundcolor=\color{lightgray},
    basicstyle=\ttfamily\scriptsize,
    breaklines=true,
    frame=none,
    showstringspaces=false,
    numbers=none,
    numberstyle=\tiny,
    literate={→}{{$\rightarrow$}}1
             {–}{{--}}1
             {“}{{``}}1
             {”}{{''}}1
}

\lstdefinelanguage{json}{
  basicstyle=\ttfamily\footnotesize,
  numbers=none,
  showstringspaces=false,
  breaklines=true,
  frame=single,
  backgroundcolor=\color[RGB]{250,250,250},
  escapeinside={(*@}{@*)},
  morestring=[b]",
  literate=
   *{:}{{{\color{black}:}}}{1}
    {,}{{{\color{black},}}}{1}
    {"}{{{\color{blue}"}}}{1}
    {[}{{{\color{black}[}}}{1}
    {]}{{{\color{black}]}}}{1}
    {\{}{{{\color{black}\{}}}{1}
    {\}}{{{\color{black}\}}}}{1},
  alsoletter={0123456789},
  stringstyle=\color{teal},
  morecomment=[l]{//},
  commentstyle=\color{gray}\ttfamily,
}

\usepackage{hyperref}
\usepackage{url}
\usepackage{cleveref}

\usepackage{amsthm}
\theoremstyle{plain}

\theoremstyle{definition}

\usepackage{comment}
\usepackage{multicol}
\usepackage{dashbox}

\begin{document}
%
\title{Towards Transparent and Incentive-Compatible Collaboration in Decentralized LLM Multi-Agent Systems: A Blockchain-Driven Approach}

\author{Minfeng Qi,~\IEEEmembership{Member,~IEEE,}
        Tianqing Zhu,~\IEEEmembership{Senior Member,~IEEE,}
        Lefeng Zhang,
        Ningran Li,
        and Wanlei Zhou,~\IEEEmembership{Fellow,~IEEE}
\thanks{M. Qi, T. Zhu, L. Zhang, W. Zhou are with the Data Science Faculty, City University of Macau, MacauSAR, China. N. Li is with the School of Computer and Mathematical Sciences, the University of Adelaide, Australia.}
\thanks{\faEnvelope{}~Corresponding author: T. Zhu, e-mail: tqzhu@cityu.edu.mo.}
}


\markboth{}{}

%



\maketitle

\begin{abstract}
Large Language Models (LLMs) have enabled the emergence of autonomous agents capable of complex reasoning, planning, and interaction. However, coordinating such agents at scale remains a fundamental challenge, particularly in decentralized environments where communication lacks transparency and agent behavior cannot be shaped through centralized incentives. We propose a blockchain-based framework that enables transparent agent registration, verifiable task allocation, and dynamic reputation tracking through smart contracts. The core of our design lies in two mechanisms: a matching score-based task allocation protocol that evaluates agents by reputation, capability match, and workload; and a behavior-shaping incentive mechanism that adjusts agent behavior via feedback on performance and reward. Our implementation integrates GPT-4 agents with Solidity contracts and demonstrates, through 50-round simulations, strong task success rates, stable utility distribution, and emergent agent specialization. The results underscore the potential for trustworthy, incentive-compatible multi-agent coordination in open environments.
\end{abstract}


%
\IEEEpeerreviewmaketitle

\section{Introduction}
\IEEEPARstart{R}ecent progress in Large Language Models (LLMs) has reshaped how we build intelligent systems. With strong abilities in reasoning, planning, and language understanding, LLMs have become the core of a new class of autonomous agents. These agents are now used in diverse areas such as scientific research, vulnerability detection, robotics, and complex problem solving~\cite{yang2025autohma,zhou2024semantic,yao2025enhancing,sun2025llm,he2025llm,wei2025advanced}.

As tasks become more demanding, multi-agent systems (MAS) built on LLMs are emerging as a promising solution for scalable intelligence. By dividing work across multiple agents, MAS can support specialization and parallel execution that help tackle challenges beyond the reach of a single agent. Popular frameworks like MetaGPT~\cite{hong2024metagpt}, AutoGen~\cite{Wu2023AutoGen}, AgentVerse~\cite{Chen2023AgentVerse}, AgentCoder~\cite{huang2023agentcoder} reflect this direction, enabling agents to plan, communicate, and coordinate. Yet, despite growing interest, a key question remains: How can we ensure that autonomous agents interact in a way that is \textit{trustworthy}, \textit{transparent}, and aligned with \textit{incentives}, especially in decentralized settings?

Many existing LLM-MAS frameworks rely on strong assumptions. They often depend on centralized control, fixed communication rules, or trusted participants~\cite{guo2024large,cinkusz2024towards,epperson2025interactive,zhang2025agentfm}. These constraints make it difficult to apply such systems in open or adversarial environments, where agents may vary in design, pursue their own goals, or behave unpredictably. In particular, current frameworks face three major limitations:

\begin{itemize}

\item \textit{Opaque collaboration mechanisms.}
Many frameworks depend on off-chain coordination, where task assignment and communication happen outside any verifiable system. This lack of transparency makes it hard to audit interactions or resolve disputes, especially in untrusted environments.

\item \textit{Missing incentive structures.}
Without mechanisms for rewards or penalties, agents lack motivation to report their capabilities honestly or complete tasks with care. This weakens collaboration, encourages poor task selection, and leads to unstable performance.

\item \textit{Limited scalability.}
Fixed protocols and centralized controllers do not scale well as agent populations grow or environments change. This creates bottlenecks in coordination and reduces the system’s ability to operate effectively in real-world, open settings.

\end{itemize}

Several recent studies have attempted to address these shortcomings through different design strategies, yet each remains limited in scope. DeCoAgent~\cite{jin2024decoagent} introduces a decentralized task marketplace, where smart contracts manage agent registration, capability discovery, and task publication. However, its design emphasizes standardization rather than dynamic incentives. BlockAgents~\cite{chen2024blockagents} takes a different direction by embedding a Proof-of-Thought consensus protocol into the collaboration workflow. While this protocol improves robustness against Byzantine behaviors, it does not fully address heterogeneous participation. LLM-Net~\cite{chong2025llmnet} proposes a blockchain-based expert network in which specialized LLM providers are selected according to immutable reputation records maintained by validators. Although this approach highlights specialization, it assumes that reputation can serve as a tamper-resistant proxy for agent quality. Taken together, these systems provide useful starting points but still fall short of simultaneously ensuring transparency, incentive alignment, and scalability in open multi-agent environments. 

\smallskip
\noindent\textbf{Our Approach.}
To address these challenges, we introduce a blockchain-based learning framework for LLM agents that supports verifiable collaboration, dynamic task allocation, and incentive-driven learning. The core idea is to use smart contracts not just for logging, but as active components that govern agent identity, task eligibility, communication, and reputation, all in a transparent and decentralized way.

Our protocol includes:
(1) cryptographic agent registration and authentication;
(2) encrypted off-chain exchanged message storage and on-chain auditability;
(3) score-matching task allocation based on reputation, capability match, and current workload;
(4) feedback-based updates to agent reputation and capability, shaped by an incentive model compatible.

To provide a clearer perspective on design characteristics and trade-offs between existing decentralized LLM-MAS systems and ours, we summarize in Table~\ref{tab:llm_mas_comparison}.

\begin{table}[htbp]
\centering
\scriptsize
\renewcommand{\arraystretch}{1.1}
\setlength{\tabcolsep}{1.6pt}
\begin{threeparttable}
\caption{Comparison of Blockchain-enabled LLM-MAS}
\label{tab:llm_mas_comparison}
\begin{tabularx}{\linewidth}{>{\raggedright\arraybackslash}m{1.9cm} 
                                    >{\centering\arraybackslash}X 
                                    >{\centering\arraybackslash}X 
                                    >{\centering\arraybackslash}X 
                                    >{\centering\arraybackslash}X}
\toprule
\textbf{Dimension} & \textbf{DeCoAgent~\cite{jin2024decoagent}} & \textbf{BlockAgents~\cite{chen2024blockagents}} & \textbf{LLM-Net~\cite{chong2025llmnet}} & \textbf{Ours} \\
\midrule
Task Decomposition      & \xmark & \xmark & \xmark & \cmark \\
Assignment Logic        & Marketplace (publish--claim) & Voting & Reputation & Utility Score \\
Agent Specialization    & \xmark & \xmark & \cmark & \cmark \\
Incentive Mechanism     & Fixed Reward & Stake-based (PoT) & Token Transfer & Behavior-Shaping \\
Behavior Tracking       & \wmark & Reasoning Audit & \wmark & On-chain + Rep. \\
Utility Modeling        & \xmark & \xmark & \xmark & \cmark \\
Performance Evolution   & \xmark & \xmark & \xmark & \cmark \\
On-chain Transparency   & \cmark & \cmark & \cmark & \cmark \\
Validation Type         & Prototype (Ethereum) & Simulations (attack benchmarks) & Simulations & Prototype + simulation\\
\bottomrule
\end{tabularx}

\vspace{0.2em}
\begin{tablenotes}[flushleft]
\footnotesize
\item Symbols: \xmark{} = Not supported, \wmark{} = Partially supported, \cmark{} = Fully supported. 
\item Abbreviations: Rep. = Reputation; PoT = Proof-of-Thought.
\end{tablenotes}
\end{threeparttable}
\end{table}

We implement the proposed framework using Solidity smart contracts, GPT-4 agents, and a React–FastAPI middleware. In a 50-round simulation, the system demonstrates consistent improvements in task completion rates, agent utility, and matching quality, while maintaining low coordination latency and stable overall performance. The experiments further reveal emerging patterns of agent specialization, indicating that the incentive mechanism fosters cooperative and adaptive behavior.

\smallskip
\noindent\textbf{Contributions.}
In summary, this work makes the following key contributions:

\ding{192} We present the first end-to-end blockchain-based LLM-MAS framework (\S\ref{sec:system}) that enables fully decentralized, transparent, and incentive-compatible multi-agent collaboration.

\ding{193} We introduce a matching score-based task allocation and a behavior-shaping incentive mechanism (\S\ref{sec:incentive-model}) that incorporates agent reputation, capability profiling, and workload balancing, along with an on-chain validation.

\ding{194} We provide an open-source prototype\footnote{open source code: \url{https://anonymous.4open.science/r/ai-agent-blockchain-collaboration-D8B2/README.md}} and extensive simulation-based evaluation (\S\ref{sec:experiment}), demonstrating the framework’s performance under decentralized environments.

\section{Preliminaries}

\subsection{Blockchain Fundamentals}

Blockchain is a distributed ledger technology that achieves data transparency, immutability, and high security through cryptographic techniques and decentralized consensus mechanisms~\cite{qi2024sok,xiao2025parallelizing}. These core characteristics make blockchain a valuable infrastructure component for enhancing transparency in traditional MAS. In our proposed framework, blockchain serves several main purposes: ensuring the transparency of task allocation, enforcing incentive protocol through smart contracts, and enabling trustworthy communication among agents (as summarized in Table~\ref{tab:bc_mas_summary}).


\begin{table}[htbp]
\centering
\caption{Comparison of Traditional MAS and Blockchain-Enhanced MAS}
\label{tab:bc_mas_summary}
\renewcommand{\arraystretch}{1.25}
\setlength{\tabcolsep}{3pt}
\begin{tabular}{m{2.4cm} m{2.9cm} m{2.9cm}}
\toprule
\textbf{Dimension} & \textbf{Traditional MAS} & \textbf{With Blockchain} \\
\midrule
Transparency           & \xmark\ Opaque & \cmark\ Globally auditable \\
Immutability           & \xmark\ Logs rewritable & \cmark\ Hash-anchored \\
Security               & \wmark\ Infra-dependent & \cmark\ Cryptographic trust \\
Task Allocation        & \xmark\ Centralized heuristic & \cmark\ Fair and decentralized \\
Incentive        & \xmark\ Lacking mechanisms & \cmark\ Tokenized rewards \\
Communication    & \xmark\ Trust-based & \cmark\ Tamper-evident \\
\bottomrule
\end{tabular}
\end{table}

\smallskip
\noindent\textbf{Blockchain data structure.}  
The blockchain is organized as a linear, tamper-resistant sequence of blocks, where each block is linked to its predecessor via a cryptographic hash of the previous block’s header~\cite{wang2025epass}. The hash of a block satisfies the following relationship:
\[
H_{\text{block}} = \text{Hash}(H_{\text{prev}}, T, M_{\text{root}}, \text{Nonce}),
\]
where $H_{\text{prev}}$ denotes the hash of the previous block, enabling the chain-like linkage; $T$ is a timestamp indicating when the block was created; $M_{\text{root}}$ is the Merkle root summarizing all transactions within the block; and \textit{Nonce} is a variable adjusted in proof-of-work (PoW) consensus to satisfy specific hash difficulty requirements. Any unauthorized alteration of block data will change $H_{\text{block}}$, thus invalidating the chain’s integrity.

Blockchain uses a Merkle tree to organize transactions within each block. A Merkle tree is a binary tree in which leaf nodes contain the hash of individual transactions, and non-leaf nodes contain hashes derived from their children:
\[
M_{\text{node}} = \text{Hash}(M_{\text{left}}, M_{\text{right}}),
\]
where $M_{\text{left}}$ and $M_{\text{right}}$ represent the hash values of the left and right child nodes, respectively. The Merkle root $M_{\text{root}}$, stored in the block header, enables secure verification of the inclusion of a specific transaction in the block. For instance, to verify whether a transaction is part of a block, one needs only to check if the computed hash path from the transaction up to the Merkle root matches the root recorded in the block header.

\smallskip
\noindent\textbf{On-chain and off-chain storage.}  
In practice, storing all data directly on-chain is inefficient and costly. Therefore, blockchain-based systems often adopt a hybrid storage model that combines on-chain and off-chain storage. Off-chain data is typically managed using distributed file systems such as the InterPlanetary File System (IPFS) or conventional databases. The integrity of off-chain data is ensured by recording its cryptographic hash on-chain~\cite{qi2022databox}. This enables verification via the following condition:
\[
\text{Verify}(H_{\text{data}}, D_{\text{off-chain}}) = 
\begin{cases} 
\text{True}, & \text{if } \text{Hash}(D_{\text{off-chain}}) = H_{\text{data}}, \\
\text{False}, & \text{otherwise.}
\end{cases}
\]
This mechanism ensures that any tampering with off-chain data can be detected by a mismatch in its hash, thereby maintaining the verifiability and trustworthiness of externally stored information.

\subsection{Multi-Agent Systems}

\begin{figure}[t]
    \centering
    \includegraphics[width=\columnwidth]{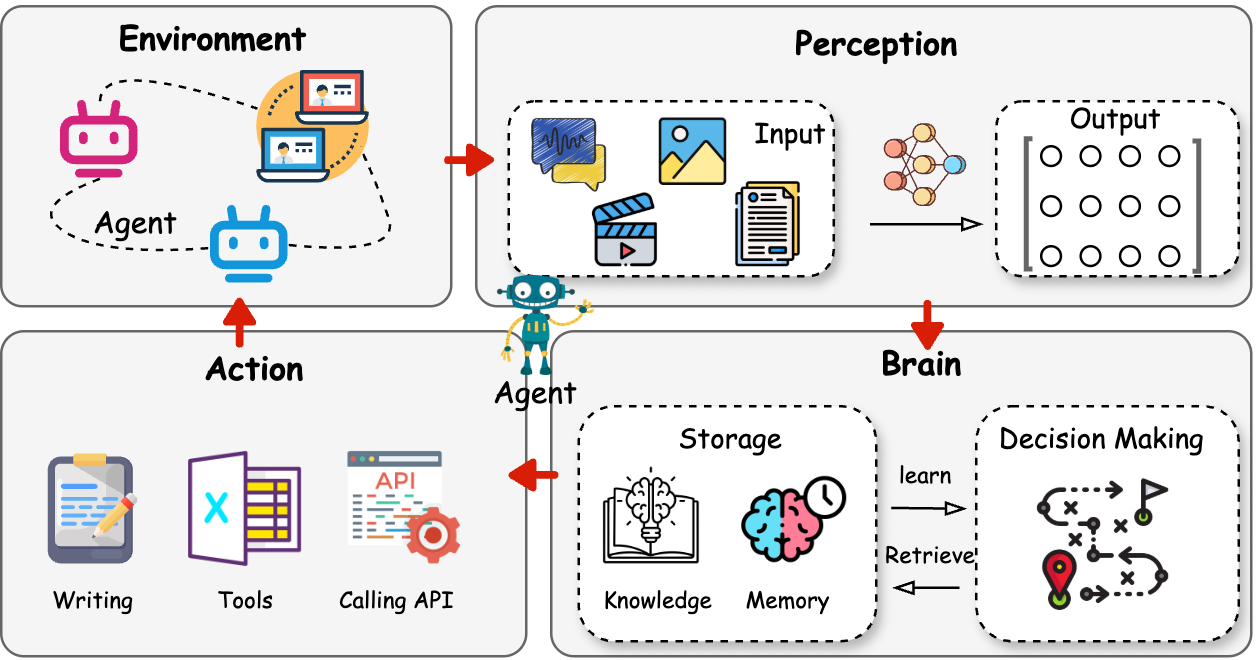}
    \caption{LLM-MAS Closed-Loop Architecture.
    The agent continuously interacts with the environment through four major components:
    (i)~\textit{Perception Module} senses raw inputs from the environment;
    (ii)~\textit{Brain Module} manages reasoning, knowledge, and planning;
    (iii)~\textit{Action Module} executes decisions that affect the environment;
    and (iv)~updated environmental states are fed back into perception, forming a closed loop.}
    \label{fig:llm-mas}
\end{figure}

MAS are composed of multiple autonomous agents that interact within a shared environment to accomplish individual or collective objectives~\cite{jiang2024multi,wang2024macrec}. These agents engage in a continuous cycle of perception, decision-making, and action, allowing them to adapt to dynamic contexts, cooperate or compete with other agents, and address complex problems beyond the capacity of a single agent (as illustrated in Fig.~\ref{fig:llm-mas}).

\smallskip
\noindent\textbf{Perception.}
Perception refers to the agent’s ability to sense and interpret environmental signals. This process transforms raw inputs (such as text queries, sensor data, or visual information) into structured internal representations that guide subsequent reasoning~\cite{pan2025robust}. In the case of LLM agents, this often involves parsing natural language input, extracting relevant entities or instructions, and mapping them into task-relevant states or semantic structures.

\smallskip
\noindent\textbf{Cognition and decision-making.}
At the core of each agent lies a reasoning module, often referred to as its cognitive engine or “brain.” This component handles information processing, memory management, contextual reasoning, and task planning~\cite{zhao2025personalized}. It stores relevant knowledge, processes new inputs in light of that knowledge, and decides on appropriate actions. Reasoning can be implemented via predefined rules, symbolic logic, or data-driven models such as reinforcement learning. LLMs in particular provide strong generalization capabilities, allowing agents to respond flexibly to previously unseen instructions.

\smallskip
\noindent\textbf{Action.}
After processing environmental input and selecting a course of action, the agent performs an action that either affects the external environment or influences other agents. This action may take various forms, including issuing API calls, generating responses, altering internal state, or sending structured messages~\cite{RenTifs2025}. The outcome of an action changes the environment, thereby initiating a new cycle of perception.

\smallskip
\noindent\textbf{Environment.}
In MAS, the environment represents the external context with which agents continuously interact~\cite{pan2025feature}. For LLM-based agents, this environment can be physical, such as a robotic workspace; digital, such as web APIs and databases; or interaction-specific, such as a user’s intent or a conversational history. The environment serves as both the source of stimuli for agent perception and the target of agent actions.

\smallskip
\noindent\textbf{Agent loop.}
This perception-decision-action process forms a continuous loop that governs the agent’s behavior. As the environment evolves, agents re-perceive, re-evaluate, and re-act, enabling adaptation to dynamic conditions and coordination with other agents~\cite{guan2025large}. This loop is particularly important in collaborative settings where agent behaviors must remain synchronized with environmental changes and peer activity.

\begin{figure*}[t]
    \centering
    \includegraphics[width=0.80\linewidth]{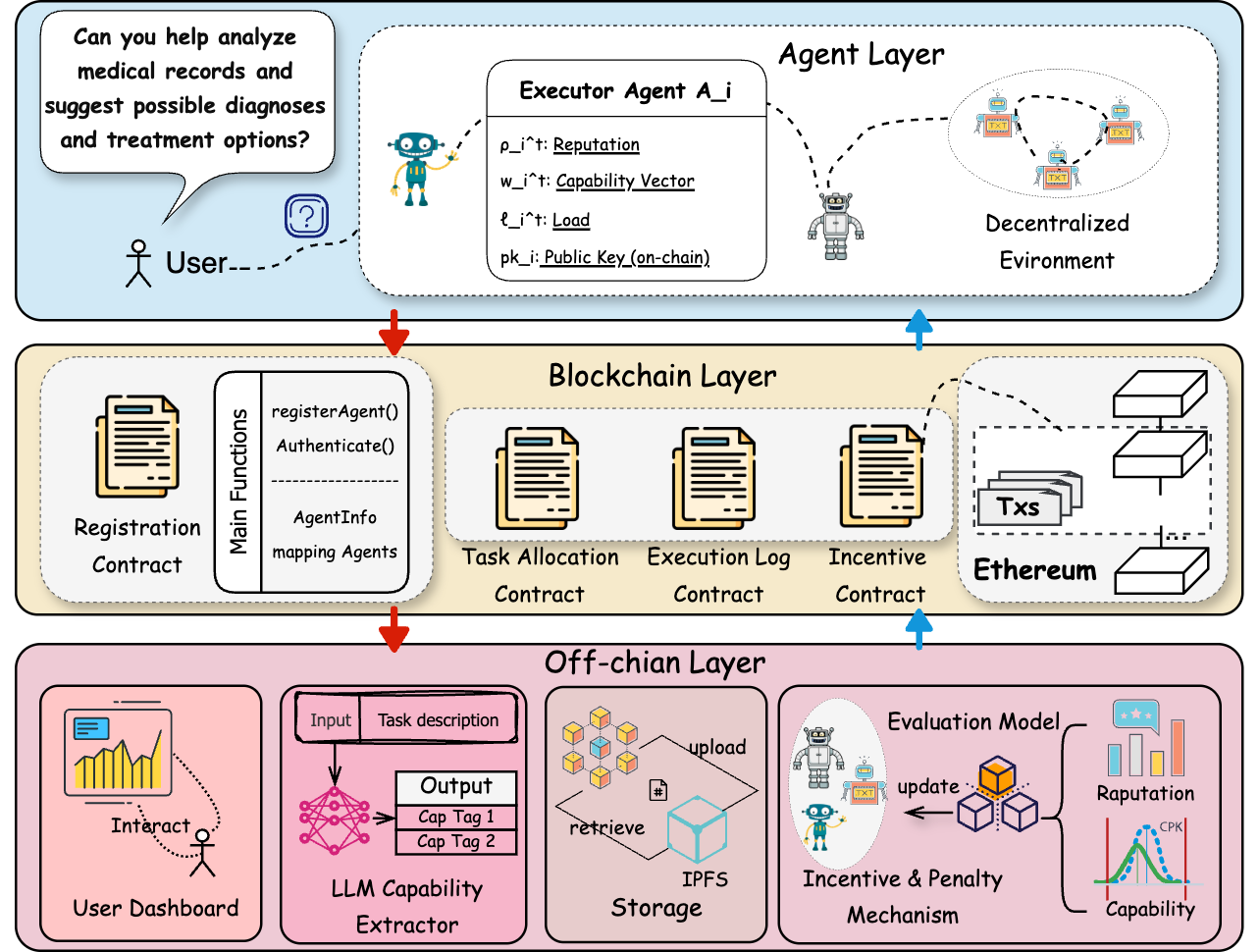}
    \caption{
        \textbf{System Architecture of Blockchain-based LLM Multi-Agent Collaboration.}
        The system is structured into three logical layers. The \textit{Off-chain Layer} handles user interactions, LLM capability extraction, storage, and performance evaluation. The \textit{Blockchain Layer} supports agent registration, task allocation, execution logging, and incentive mechanisms via smart contracts deployed on Ethereum. The \textit{Agent Layer} consists of autonomous agents operating in a decentralized environment, each maintaining reputation, capability, load, and public key information.
    }
    \label{fig:architecture}
\end{figure*}

\section{System Design}
\label{sec:system}
To enable transparent and autonomous coordination among LLM-based agents, we propose a smart contract-driven interaction protocol for multi-agent systems (cf.~Fig.~\ref{fig:architecture}).

The protocol consists of four core modules: (1) agent registration and authentication, (2) communication and messaging, (3) decentralized task allocation, and (4) incentive management. Each module is implemented through dedicated smart contracts and interacts with both on-chain and off-chain components to support verifiable, adaptive collaboration among agents.

\subsection{Registration and Authentication Module}
The core functionality of this module includes agent identity registration and subsequent identity authentication during interactions. Through this module, each agent obtains a unique and verifiable digital identity, which serves as a foundational guarantee for ensuring transparency and trustworthiness in the operation of the multi-agent system.

\smallskip
\noindent\textbf{Registration phase.} In the registration phase, each agent independently generates a public-private key pair on its local device to establish a unique cryptographic identity. This decentralized key management approach ensures that private keys never leave the agent's trusted environment, thereby enhancing privacy, preventing single points of failure, and aligning with trust-minimized system design principles.

To ensure interoperability and cryptographic uniformity, the protocol mandates a standardized key scheme. Specifically, all agents are required to use the $\mathsf{secp256k1}$ elliptic curve digital signature algorithm, which is widely adopted in blockchain systems such as Bitcoin and Ethereum. The public key is encoded in compressed form and submitted to the blockchain during registration.

Each agent completes registration by invoking the $\mathsf{registerAgent}$ function exposed by the smart contract, providing the following fields as input:

\begin{itemize}
    \item \textit{AgentID}: A unique identifier for the agent, used to index agent state and enable interaction.
    \item \textit{PublicKey}: The agent’s self-generated $\mathsf{secp256k1}$ public key, used for signature verification.
    \item \textit{Role}: A declared functional role (e.g., $\mathsf{Executor, Issuer}$), which governs the agent’s permitted operations.
    \item \textit{Capabilities}: A set of capability descriptors such as “NLP Analysis” or “Data Classification”, which are used to match agents with appropriate tasks.
    \item \textit{Reputation}: An optional initial reputation score, which may be externally verified or initialized with a system default.
\end{itemize}

Upon receiving the registration request, the contract validates that (i) the caller address is not already registered, (ii) mandatory fields are non-empty, and (iii) the public key complies with the expected format and cryptographic curve. If validation passes, the agent's metadata is stored in a mapping $\mathsf{agents}$ indexed by blockchain address. The mapping holds the agent’s identity attributes, cryptographic credentials, role, capability tags, reputation score, and other operational metadata.

The smart contract then sets the $\mathsf{isRegistered}$ flag to $\mathsf{true}$ and emits an $\mathsf{AgentRegistered}$ event containing the registration details. This event enables transparent indexing and monitoring by external services or observer contracts. Additionally, the agent's task load counter is initialized to zero, serving as a baseline for workload-aware task allocation.

\smallskip
\noindent\textbf{Authentication phase.}
In order to ensure that only registered and qualified agents are permitted to participate in subsequent interactions, our protocol includes a robust authentication mechanism based on digital signature verification. This process enables any agent to cryptographically prove its identity without revealing its private key. The authentication workflow proceeds as follows.

First, the agent generates a challenge message locally (e.g., $\mathsf{"Authenticate\ me"}$) and signs it using its private key. The resulting signature, along with the original message, is submitted to the blockchain by invoking the $\mathsf{authenticate}$ function of the smart contract.

Upon receiving the authentication request, the contract performs several cryptographic checks to validate the agent's identity. Specifically, it computes the Keccak-256 hash of the original message, yielding a $\mathsf{messageHash}$. This hash is then prefixed according to Ethereum’s standard signing scheme to produce an $\mathsf{ethSignedMessageHash}$. The contract invokes the $\mathsf{ecrecover}$ function on the hash and the submitted signature to recover the Ethereum address of the signer. If the recovered address matches the address of the caller, and the caller is already registered in the $\mathsf{agents}$ mapping, the authentication is considered successful.

This signature-based challenge-response protocol ensures that agents cannot spoof identities or forge interactions, as only the legitimate holder of the corresponding private key can produce a valid signature.

In the context of task execution, additional capability-based validation is also enforced. When an agent requests to apply for or execute a task, the smart contract invokes the $\mathsf{verifyCapabilities}$ function to check whether the agent possesses all the required skills for the task. This is performed by comparing the set of capability tags stored during registration with the set of tags associated with the task. If the agent’s capability set fully satisfies the task’s requirement set, the request is accepted; otherwise, the contract rejects the operation.

\subsection{Communication and Messaging Module}

In a MAS, communication and message passing are fundamental to achieving coordinated behavior among agents~\cite{hu2024learning,ding2024learning}. This module is designed to ensure that inter-agent communication is secure, transparent, and efficient, while maintaining an immutable record of all message exchanges for future traceability. By leveraging the tamper-resistant nature of blockchain and the automation capabilities of smart contracts, the communication and messaging module implements decentralized message management and verification.

\smallskip
\noindent\textbf{Message structure design.}
To ensure consistency and verifiability of inter-agent communication, all messages must adhere to a standardized structure comprising the following fields: sender, receiver, timestamp, message type, message body, and digital signature. 

The \textit{sender} and \textit{receiver} fields identify the parties involved in the communication, ensuring that messages are properly addressed. The \textit{timestamp} records the time of message creation, which is critical for preventing replay attacks and maintaining event ordering. The \textit{message\_type} indicates the functional intent of the message (e.g., task assignment, status update, or result feedback), guiding the receiving agent on how to process it. The \textit{message\_body} contains the actual payload, such as task descriptions, execution state, or contextual metadata. Finally, the \textit{signature} is generated by the sender using its private key, enabling the receiver and other observers to verify the message's authenticity and integrity.

\begin{lstlisting}[language=json, caption={Standardized Message Structure}, label={lst:message-structure}]
{
  "message_id": "UUID",
  "sender": "Agent_A",
  "receiver": "Agent_B",
  "timestamp": "2025-06-12 10:30:00",
  "message_type": "TaskAssignment",
  "task_id": "Task_UUID",
  "capability_tags": ["Tag1", "Tag2"],
  "status": "InProgress",
  "message_body": {
    "description": "Task or status update description",
    "data": "Associated data or context"
  },
  "signature": "digital_signature"
}
\end{lstlisting}

The use of a unified message schema facilitates automatic parsing, protocol enforcement, and formal verification. Signatures are verified on-chain or off-chain using the agent’s registered public key, ensuring that only authenticated agents can issue valid messages. The inclusion of task and capability metadata also supports downstream validation steps, such as task eligibility checks and capability matching.

To balance performance and transparency, the system supports hybrid logging: full message content may be stored off-chain (e.g., in IPFS or a decentralized database), while the message hash and metadata are anchored on-chain. This hybrid architecture significantly reduces storage overhead without sacrificing data integrity or auditability.

\begin{figure}[t]
    \centering
    \includegraphics[width=\columnwidth]{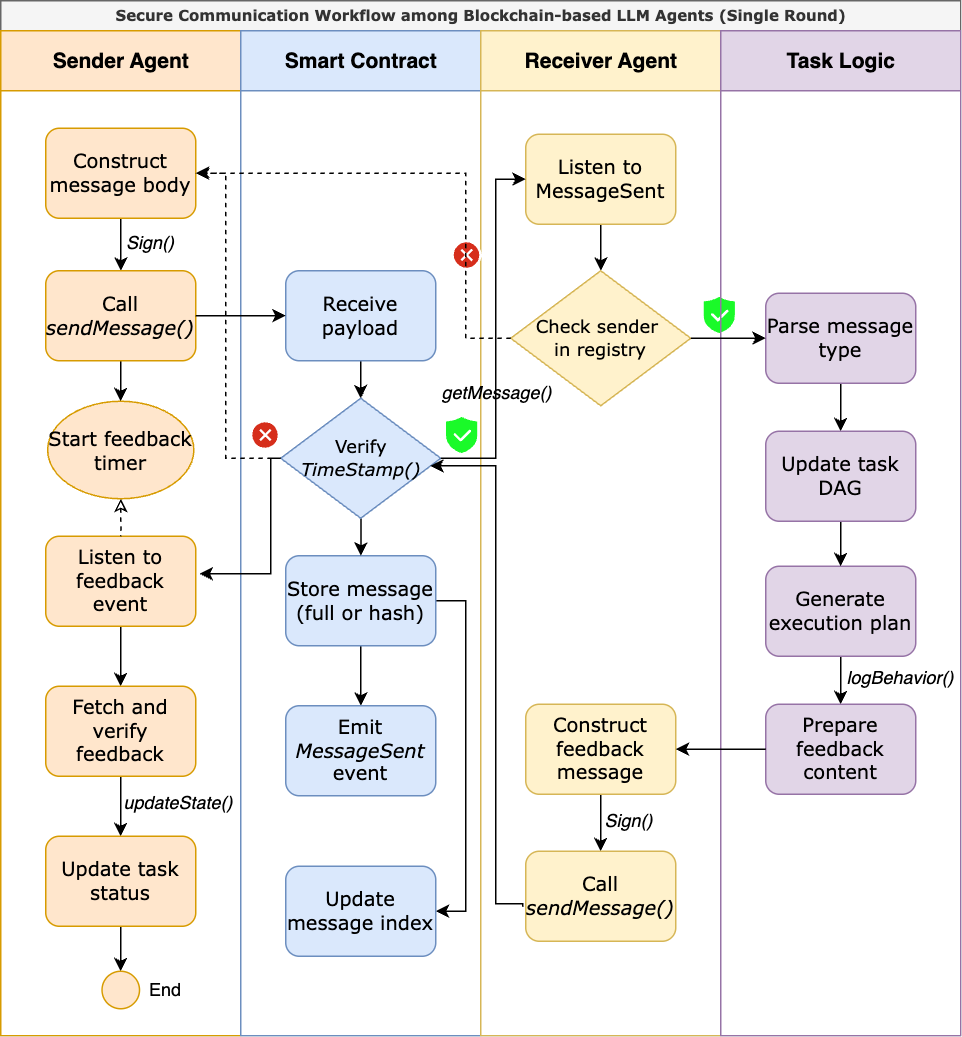}
    \caption{End-to-end message exchange process across four logical components: 
    (i)~\textit{Sender Agent} constructs and signs the message; 
    (ii)~\textit{Smart Contract} verifies and records it with a timestamp; 
    (iii)~\textit{Receiver Agent} validates the sender and generates feedback;
    and (iv)~\textit{Task Logic} parses the message and updates task state.
    Feedback is looped back securely through the same contract-based mechanism.}
    \label{fig:llm-agent-communication}
\end{figure}

\smallskip
\noindent\textbf{Communication workflow.}
The communication workflow defines the lifecycle of message exchange between agents, encompassing message construction, on-chain transmission, cryptographic verification, task-specific processing, and optional feedback generation (as presented in Fig.~\ref{fig:llm-agent-communication}).

\noindent\textit{(1) Message generation.}  
To initiate communication, the sending agent first constructs the message body based on the interaction context. This may include task descriptions, intermediate execution states, or final results, depending on the communication objective. The sender then populates metadata fields such as $\mathsf{sender}$, $\mathsf{receiver}$, $\mathsf{timestamp}$, and $\mathsf{message\_type}$. These metadata elements provide critical context for the receiver and help ensure the uniqueness and traceability of the message. Once the message structure is complete, the sender uses its private key to sign the message content—comprising both metadata and body—thereby generating a $\mathsf{signature}$ field. This digital signature guarantees message integrity and non-repudiation, preventing tampering and impersonation during transmission.

\noindent\textit{(2) Message transmission.}  
The constructed message is submitted to the blockchain by invoking the $\mathsf{sendMessage}$ function of the smart contract. Upon receiving the message, the contract first verifies the $\mathsf{timestamp}$ to ensure that the message is fresh (e.g., generated within the last 5 minutes), thereby mitigating replay attacks. Once the timestamp is validated, the message is recorded in the on-chain message log. For high-priority messages (e.g., task requests or completion reports), the full content is stored on-chain. For lower-priority messages (e.g., routine status updates), only the message hash may be stored to reduce blockchain storage costs. After recording the message, the contract emits a $\mathsf{MessageSent}$ event containing the sender address, receiver address, and message type. The receiving agent listens for this event to detect the arrival of new messages and trigger the verification process.

\noindent\textit{(3) Message verification.}  
Upon receiving a notification, the receiver fetches the message content from the smart contract and initiates a verification procedure to ensure authenticity and integrity. The first step involves checking the $\mathsf{timestamp}$ to confirm that the message is still within the valid time window, defending against delayed or replayed transmissions. The receiver then invokes the $\mathsf{verifyMessage}$ function provided by the contract. This verification includes computing the $\mathsf{messageHash}$ from the content, converting it to the Ethereum standard format ($\mathsf{ethSignedMessageHash}$), and applying the $\mathsf{ecrecover}$ function to extract the sender's address from the signature. If the recovered address matches the $\mathsf{sender}$ field in the message and corresponds to a registered agent, the message is considered authentic and intact. Only successfully verified messages proceed to execution.

\noindent\textit{(4) Message handling.}  
Once validated, the message is dispatched to the receiver’s task logic according to its $\mathsf{message\_type}$. For example, in the case of a task assignment message, the receiver parses the task description, evaluates feasibility, and decides whether to accept the task. If accepted, the agent may generate an execution plan and record the task's initial state. For a status update message, the agent interprets the updated parameters and adjusts its local task scheduler or planning module accordingly. If the message carries a result report, the agent logs the outcome and may forward it to upstream coordinators or integrate it into downstream decisions.

\noindent\textit{(5) Feedback generation.}  
In some interaction scenarios, the receiving agent is required to issue a response message containing the result of processing. This involves generating a new message with updated content, including processing results, metadata (e.g., sender and receiver addresses, timestamps), and a fresh digital signature. The receiver then calls the $\mathsf{sendMessage}$ function to deliver the response back to the blockchain. As with previous transmissions, the message is logged, and a new $\mathsf{MessageSent}$ event is emitted to notify the original sender. The initial sender listens for this event, retrieves the feedback, and performs signature verification and semantic interpretation. 

\subsection{Task Allocation Module}

Task allocation in our multi-agent framework is governed by a dynamic policy that balances agent capabilities, reputation, and current workload. At the core of this process lies the task assignment policy \( \pi_i^t(T_j) \in [0, 1] \), which denotes the probability that agent \( A_i \) is assigned to task \( T_j \) at time \( t \). This probability is computed based on a utility-driven scoring function that supports rational agent selection and incentive alignment, ensuring that high-quality agents are consistently matched to appropriate tasks.

\noindent\textit{(1) Task description and capability tagging.}
Each task \( T_j \) is described in natural language by the task issuer and then parsed using an LLM-based capability extractor. The system transforms the task into a structured representation containing a set of required capabilities \( \mathcal{C}_j^{\text{req}} \subseteq \mathcal{T} \), where \( \mathcal{T} \) is the global capability tag space. This set reflects the domain-specific knowledge, technical operations, and linguistic tasks involved.

\noindent\textit{(2) Agent capability profiling.}
Each agent \( A_i \) maintains a dynamic capability profile \( \boldsymbol{w}_i^t = (w_{i,1}^t, \ldots, w_{i,K}^t) \in [0,1]^K \), which quantifies its current proficiency across \( K \) predefined capabilities. Alongside this, the agent’s reputation score \( \rho_i^t \in [0,1] \) summarizes its historical performance, while \( \ell_i^t \in \mathbb{N} \) denotes its current task load.

\noindent\textit{(3) Matching score and assignment probability.}
To compute \( \pi_i^t(T_j) \), we first define a scoring function:
\begin{align}
\label{scoring}
    S_{i,j}^t = \lambda_1 \cdot \mathsf{CapMatch}(\mathcal{C}_j^{\text{req}}, \boldsymbol{w}_i^t) + \lambda_2 \cdot \rho_i^t - \lambda_3 \cdot \ell_i^t
\end{align}
Here, \( \mathsf{CapMatch}(\mathcal{C}_j^{\text{req}}, \boldsymbol{w}_i^t) = \frac{1}{|\mathcal{C}_j^{\text{req}}|} \sum_{k \in \mathcal{C}_j^{\text{req}}} w_{i,k}^t \) represents the average capability weight of agent \( A_i \) on the required tags. The weights \( \lambda_1, \lambda_2, \lambda_3 \geq 0 \) determine the importance of each component.

The task assignment probability is then computed via a softmax distribution over candidate agents:
\begin{align}
\label{task-assignment}
    \pi_i^t(T_j) = \frac{\exp(S_{i,j}^t)}{\sum_{A_k \in \mathcal{A}_j} \exp(S_{k,j}^t)}
\end{align}
where \( \mathcal{A}_j \) is the set of eligible agents for task \( T_j \). This formulation ensures that higher-scoring agents are more likely to be selected while still allowing exploration and diversity.

\noindent\textit{(4) Collaborative task decomposition.} 
For tasks with a complexity level that exceeds a threshold \( \theta_c \), the task is decomposed into subtasks \( \{T_j^{(1)}, \ldots, T_j^{(m)}\} \). Each subtask inherits a subset of capability requirements, and the above assignment policy \( \pi_i^t(T_j^{(k)}) \) is applied independently. The smart contract enforces dependency constraints and execution order, facilitating multi-agent collaboration.

\noindent\textit{(5) Post-assignment update.} 
Upon assignment, the selected agent's load \( \ell_i^t \) is incremented by one, and task state is updated on-chain. After task completion and evaluation, the agent’s reputation \( \rho_i^{t+1} \) and capability vector \( \boldsymbol{w}_i^{t+1} \) are updated based on the task score, as described in the incentive mechanism section (\S\ref{sec:incentive-model}). This feedback loop ensures that future task allocation decisions are continuously adapted to agent performance.

\section{Behavior-shaping Incentive Mechanism}
\label{sec:incentive-model}

As a central innovation of our framework, we design a behavior-shaping mechanism (Alogrithm~\ref{alg:incentive-mechanism}) that couples short-term utility optimization with long-term trust shaping. Specifically, agent behavior is dynamically regulated through a protocol that (i) computes utilities based on reward, workload, and capability mismatch, (ii) updates reputation scores via performance-sensitive smoothing, and (iii) adjusts capability weights according to task-specific outcomes. By integrating these components, the mechanism not only rewards high-quality task execution but also penalizes poor performance, thereby steering agents toward cooperative behavior over interactions.

\begin{algorithm}[htbp]
\caption{Behavior-shaping Incentive Mechanism for Decentralized Agents Coordination}
\label{alg:incentive-mechanism}
\KwIn{Task set $\mathcal{T}$, Agent set $\mathcal{A}$, capability taxonomy $\mathcal{T}_{\text{cap}}$}
\KwOut{Updated $\rho_i^{t+1}$ and $\boldsymbol{w}_i^{t+1}$ for all $A_i \in \mathcal{A}$}

\ForEach{task $T_j \in \mathcal{T}$}{
    Retrieve reward $R_j$, required tag vector $\mathbf{r}_j$, and deadline $\tau_j$\;

    \ForEach{agent $A_i \in \mathcal{A}$}{
        Compute binarized capability $\hat{\mathbf{w}}_i^t$ from $\boldsymbol{w}_i^t$ using threshold $\theta$\;

        Compute cost: $c_i^t(T_j) = \beta \cdot \ell_i^t + \gamma \cdot \left\| \mathbf{r}_j - \hat{\mathbf{w}}_i^t \right\|_1$\;

        Compute utility: $U_i^t(T_j) = \pi_i^t(T_j) \cdot R_j - c_i^t(T_j)$\;

        Compute assignment probability $\pi_i^t(T_j)$ using scoring function based on $\rho_i^t$, $\boldsymbol{w}_i^t$, and $\ell_i^t$\;
    }

    Assign task $T_j$ to agent $A_{i^*}$ with highest $\pi_{i^*}^t(T_j)$ and log assignment\;
    
    Execute $T_j$ and collect performance metrics: $q_{i^*}^t$, $d_{i^*}^t$, and tag-wise $s_{i^*,k}^t$\;
    
    Compute task score: $S_{i^*}^t = \alpha \cdot q_{i^*}^t + \delta \cdot (1 - d_{i^*}^t)$\;

    Update reputation:
    $\rho_{i^*}^{t+1} = \lambda \cdot \rho_{i^*}^t + (1 - \lambda) \cdot S_{i^*}^t$\;

    \ForEach{tag $\mathsf{Tag}_k \in \mathcal{T}_j$}{
        Update capability weight:
        $w_{i^*,k}^{t+1} = \mu \cdot w_{i^*,k}^t + (1 - \mu) \cdot s_{i^*,k}^t$\;
    }

    Set $\ell_{i^*}^{t+1} \gets \ell_{i^*}^t - 1$\;
}
\end{algorithm}
\vspace{-0.2in}

\subsection{Utility-based Incentive Function}

In our framework, each participating entity is modeled as an autonomous agent operating in a decentralized environment. Let \( \mathcal{A} = \{A_1, A_2, \dots, A_n\} \) denote the set of all agents. Each agent \( A_i \in \mathcal{A} \) is characterized at time step \( t \) by a profile consisting of three core components: its reputation score \( \rho_i^t \in [0, 1] \), a capability vector \( \mathbf{w}_i^t = (w_{i,1}, \dots, w_{i,d}) \in \mathbb{R}_{\geq 0}^d \), and a workload level \( \ell_i^t \in \mathbb{N} \). The reputation score reflects the agent’s historical performance quality, the capability vector encodes proficiency across \( d \) skill domains, and the workload level records the number of active or pending tasks assigned to the agent.

Tasks are periodically issued into the environment. Each task \( T_j \) is defined by a reward value \( R_j > 0 \), a required capability vector \( \mathbf{r}_j \in \{0,1\}^d \), and an expected execution window or deadline \( \tau_j \). The binary capability vector \( \mathbf{r}_j \) specifies which skill dimensions are required for completing the task, i.e., \( r_{j,k} = 1 \) indicates that skill \( k \) is necessary.

To allocate tasks effectively, the system defines a task assignment policy \( \pi_i^t(T_j) \in [0, 1] \) (Formula \ref{task-assignment}), which represents the probability of assigning task \( T_j \) to agent \( A_i \) at time \( t \). This assignment probability is derived from the scoring function (Formula \ref{scoring}) that considers the agent’s reputation, capability match, and current workload.

The expected utility that agent \( A_i \) receives at time \( t \) when being evaluated for task \( T_j \) is defined as:
\begin{equation}
\label{equation:utility}
    U_i^t(T_j) = \pi_i^t(T_j) \cdot R_j - c_i^t(T_j),
\end{equation}

where \( \pi_i^t(T_j) \cdot R_j \) denotes the expected reward the agent may earn, and \( c_i^t(T_j) \) denotes the cost of executing the task. This cost function encapsulates both the agent’s current workload and the degree of mismatch between its skills and the task requirements. Formally, we define:
\begin{equation}
\label{equation:cost}
c_i^t(T_j) = \beta \cdot \ell_i^t + \gamma \cdot \left\| \mathbf{r}_j - \hat{\mathbf{w}}_i^t \right\|_1,
\end{equation}
where \( \beta > 0 \) and \( \gamma > 0 \) are system-defined coefficients, \( \ell_i^t \) is the current workload, and \( \hat{\mathbf{w}}_i^t \in \{0,1\}^d \) is the binarized capability vector derived from \( \mathbf{w}_i^t \) by thresholding, i.e., \( \hat{w}_{i,k}^t = 1 \) if \( w_{i,k}^t \geq \theta \), and 0 otherwise. The \( L_1 \) norm \( \left\| \mathbf{r}_j - \hat{\mathbf{w}}_i^t \right\|_1 \) quantifies the number of required skills that the agent lacks.

Thus, based on Formula \ref{equation:cost} and \ref{equation:utility}, the agent’s utility function becomes:
\begin{equation}
\label{equation:utility-full}
U_i^t(T_j) = \pi_i^t(T_j) \cdot R_j - \beta \cdot \ell_i^t - \gamma \cdot \left\| \mathbf{r}_j - \hat{\mathbf{w}}_i^t \right\|_1.
\end{equation}

This formulation implies that an agent's incentive to accept a task increases with a higher assignment probability \( \pi_i^t(T_j) \), larger reward \( R_j \), and better capability alignment with the task (i.e., a smaller skill mismatch). In contrast, agents are discouraged from accepting tasks when they are overloaded (\( \ell_i^t \) is large) or underqualified (i.e., many components of \( \mathbf{r}_j \) are not satisfied by \( \hat{\mathbf{w}}_i^t \)).

The incentive structure is thus designed to promote truthful reporting of capabilities and discourage strategic misrepresentation. Since utility is explicitly penalized by capability mismatch and task overload, agents have no rational incentive to overstate their abilities or accept tasks beyond their capacity. Moreover, as will be demonstrated in subsequent sections, this utility function integrates naturally with our dynamic reputation update rule and capability-weight learning mechanism, forming a self-regulating incentive structure that rewards cooperative and capable behavior over repeated interactions.

\subsection{Reputation Update Rule}

To incentivize high-quality behavior and maintain long-term agent accountability, our framework integrates a dynamic reputation update mechanism. Each agent \( A_i \in \mathcal{A} \) maintains a scalar reputation score \( \rho_i^t \in [0,1] \), which is updated over time based on its observed performance across assigned tasks. This reputation score plays a central role in influencing future task assignment probabilities and serves as a trust indicator in agent selection procedures.

Let \( S_i^t \in [0,1] \) denote the performance score that agent \( A_i \) receives upon completing a task at time step \( t \). This score is computed based on the task outcome, and can be determined either through automated metrics (e.g., accuracy, latency) or feedback from the task issuer. To ensure stability and gradual adaptation, we define the reputation update rule using an exponential moving average (EMA) as follows:
\begin{equation}
    \rho_i^{t+1} = \lambda \cdot \rho_i^t + (1 - \lambda) \cdot S_i^t,
\end{equation}

where \( \lambda \in [0,1) \) is a smoothing coefficient that determines how strongly past reputation influences future values. A higher value of \( \lambda \) yields more inertia (slower change), while a lower \( \lambda \) allows faster responsiveness to recent performance.

This update rule ensures that an agent’s reputation gradually converges toward its average task performance. Since \( S_i^t \in [0,1] \) and \( \rho_i^t \in [0,1] \), the update equation preserves boundedness: \( \rho_i^{t+1} \in [0,1] \) for all \( t \). The EMA design naturally implements a forgetting mechanism, allowing agents to recover from past errors over time while still penalizing consistently poor behavior.

The performance score \( S_i^t \) is computed as a weighted aggregation of multiple task-specific indicators:
\begin{equation}
    S_i^t = \alpha \cdot q_i^t + \delta \cdot (1 - d_i^t),
\end{equation}

where \( q_i^t \in [0,1] \) is the task quality score (e.g., correctness or user rating), \( d_i^t \in [0,1] \) is a normalized delay ratio (actual time divided by deadline), and \( \alpha + \delta = 1 \) are weighting coefficients. This formulation penalizes delayed or substandard executions and rewards timely, high-quality completions.

A key property of the reputation rule is that it aligns with the agent’s utility function \( U_i^t \) (Formula \ref{equation:utility-full}). Since task allocation probabilities \( \pi_i^t \) are influenced by \( \rho_i^t \), maximizing long-term utility requires agents to improve their reputation. Let the expected cumulative utility over a horizon \( T \) be denoted by:
\begin{equation}
    \mathbb{E}[\mathcal{U}_i] = \sum_{t=1}^T \mathbb{E}[U_i^t(T_j)].
\end{equation}

Substituting the utility expression from the previous section and noting that \( \pi_i^t \propto \rho_i^t \), it becomes clear that maximizing \( \mathbb{E}[\mathcal{U}_i] \) requires maintaining a high \( \rho_i^t \) across \( t \). This creates a natural incentive for rational agents to pursue good behavior over time.

\subsection{Capability Weight Adjustment}

In addition to reputation, each agent \( A_i \in \mathcal{A} \) maintains a dynamic capability vector \( \boldsymbol{w}_i^t \in [0,1]^K \), where each element \( w_{i,k}^t \) quantifies the agent’s expertise or reliability in a specific skill dimension \( k \in \{1, 2, \dots, K\} \). These capability weights serve as fine-grained indicators that complement the scalar reputation score and facilitate task-specific matching in heterogeneous multi-agent environments.

Let \( \mathsf{Tag}_k \) denote the \( k \)-th capability tag in the global capability taxonomy \( \mathcal{T} = \{\mathsf{Tag}_1, \mathsf{Tag}_2, \dots, \mathsf{Tag}_K\} \). The value \( w_{i,k}^t \in [0,1] \) indicates the proficiency level of agent \( A_i \) on tag \( \mathsf{Tag}_k \) at time \( t \). Initially, each agent declares a self-assessed capability vector \( \boldsymbol{w}_i^0 \) during registration. However, the system dynamically updates these values based on the agent’s historical task performance in each capability domain.

For each completed task \( T_j \) with required tag set \( \mathcal{T}_j \subseteq \mathcal{T} \), the agent receives a capability-specific performance score \( s_{i,k}^t \in [0,1] \) for each tag \( \mathsf{Tag}_k \in \mathcal{T}_j \). This local score reflects how well the agent performed the sub-tasks associated with that capability dimension. The weight update rule is again modeled using exponential smoothing:
\begin{equation}
    w_{i,k}^{t+1} = \mu \cdot w_{i,k}^t + (1 - \mu) \cdot s_{i,k}^t,
\quad \text{for all } \mathsf{Tag}_k \in \mathcal{T}_j,
\end{equation}

where \( \mu \in [0,1) \) is the capability parameter that governs the sensitivity to new observations. Tags that are not involved in the task (\( \mathsf{Tag}_k \notin \mathcal{T}_j \)) retain their previous weights: \( w_{i,k}^{t+1} = w_{i,k}^t \).

The use of tag-specific smoothing allows agents to improve (or degrade) their perceived skill levels over time in a way that is tightly aligned with actual task performance. Agents who consistently perform well in certain domains will see corresponding capability weights increase, making them more likely to be selected for similar tasks in the future. Conversely, agents with poor or unstable performance in a domain will experience weight decay, reducing their assignment likelihood for those tags.

\subsection{On-Chain Enforcement via Smart Contracts}

To ensure verifiability and autonomy of coordination among agents, our framework leverages on-chain enforcement through smart contracts. All state transitions, including agent registration, task assignment, behavioral logging, performance evaluation, and reputation updates, are conducted on-chain to benefit from blockchain’s immutability and auditability.

Each smart contract maintains the relevant state variables and enforces lifecycle transitions via publicly verifiable transactions. When an agent $A_i$ joins the system, its declared capabilities $\mathcal{C}_i$, public key, and initial reputation $r_i^0$ are stored through the invocation of the $\mathsf{registerAgent}$ function. All agent-specific operations thereafter, such as authentication, workload updates, and reputation adjustments, are tracked and enforced through designated contract interfaces.

Tasks $T_j$ are registered on-chain via $\mathsf{submitTask}$, which stores the task’s metadata, including capability requirements $\mathsf{requiredTags}_j$ and execution constraints. The assignment policy $\pi_i^t(T_j)$ is computed on-chain by evaluating agent scores based on capability match, current load, and historical reputation. The selected agent $A_i$ is recorded through a state update and confirmed by emitting a $\mathsf{TaskAssigned}$ event.

During task execution, all behavioral evidence, including task acceptance, intermediate updates, and result submission, is reported using the $\mathsf{logAction}$ interface. Each log entry binds a specific action to an agent-task pair, forming a verifiable tuple $\langle A_i, T_j, \mathsf{action}, t \rangle$. These logs serve as audit trails.

Upon task completion, performance evaluation and reputation updates are enforced on-chain via the $\mathsf{updateReputation}$ function. The function takes the task score $\mathsf{TS}_{i,j}$ as input and computes the updated reputation $r_i^{t+1}$ using a predefined rule. This ensures that all reputation transitions are transparent and irrevocable.

\section{Experimental Setup}
\label{sec:experiment}

\subsection{System Implementation}
Our system consists of three main components: (1) the LLM-based agent layer, (2) the smart contract coordination layer, and (3) the user-facing frontend. 

\noindent\textit{(1) Agent layer.} Each agent is instantiated as an independent process equipped with access to OpenAI's GPT-4 API. Agents communicate via signed messages (using ECDSA over secp256k1) and locally maintain their reputation scores, capability profiles, and decision-making modules.

\noindent\textit{(2) Smart contract layer.} The blockchain backend is implemented using the Ethereum Virtual Machine (EVM). All contract logic, including agent registration, task assignment, and reputation updates, is encoded in Solidity. We used Hardhat for development and deployed contracts on a local Ganache testnet to simulate realistic blockchain execution.

\noindent\textit{(3) Frontend and middleware.} We implemented a lightweight React-based frontend that allows users to submit tasks, monitor agent performance, and visualize blockchain logs. A FastAPI-based backend mediates between agent behavior and smart contract calls, including transaction signing, gas estimation, and off-chain capability evaluation.

The entire system was developed and tested on a machine with the following specifications:

\begin{table}[htbp]
\centering
\caption{Experimental Environment Configuration}
\label{tab:system_config}
\renewcommand{\arraystretch}{1.25}
\setlength{\tabcolsep}{4pt}
\begin{tabular}{m{3.2cm} m{5.1cm}}
\toprule
\textbf{Component} & \textbf{Specification} \\
\midrule
CPU & 12-core AMD Ryzen 9 5900X @ 3.7GHz \\
RAM & 32 GB DDR4 \\
Operating System & Ubuntu 22.04 LTS \\
Blockchain Environment & Ganache CLI v7.8.0, Solidity v0.8.20, Hardhat v2.19 \\
LLM API & OpenAI GPT-4 API (temperature = 0.2, max tokens = 1024) \\
Libraries & Web3.py, FastAPI, IPFS CLI, Numpy, Matplotlib \\
\bottomrule
\end{tabular}
\end{table}
\vspace{-0.2in}

\subsection{Simulated Task Environment}

\noindent\textbf{Task corpus and capability taxonomy.}
We derive our task set from a modified subset of the ALFRED benchmark~\cite{shridhar2020alfred}, a well-known environment used in multi-agent learning and embodied task planning. ALFRED provides hierarchical natural language instructions grounded in realistic household environments, which are particularly suitable for simulating multi-skill task execution in distributed agent systems. For our purposes, we extract 100 representative tasks involving both atomic goals (e.g., “pick up the book”) and composite objectives (e.g., “put the red mug on the kitchen table and turn off the lights”).

Each task is automatically annotated with 2–4 relevant capability tags drawn from a curated global taxonomy \( \mathcal{T} = \{\mathsf{Tag}_1, \mathsf{Tag}_2, \dots, \mathsf{Tag}_{10}\} \). This taxonomy captures diverse functional dimensions and serves as the semantic interface between task descriptions and agent skill profiles (as presented in Table~\ref{tab:capability_tags}).

Each task \( T_j \) is thus encoded as a binary capability requirement vector \( \mathbf{r}_j \in \{0,1\}^{10} \), where each component \( r_{j,k} = 1 \) indicates that the task demands skill \( \mathsf{Tag}_k \). For example, a composite instruction such as “pick up the knife and place it on the stove” may require \( \mathsf{Tag}_1 \) (object recognition), \( \mathsf{Tag}_4 \) (manipulation), and \( \mathsf{Tag}_5 \) (navigation). 

\begin{table}[htbp]
\centering
\caption{Capability Tag Taxonomy Used for Agent Profiling}
\label{tab:capability_tags}
\renewcommand{\arraystretch}{1.25}
\setlength{\tabcolsep}{4pt}
\begin{tabular}{m{1.8cm} m{6.3cm}}
\toprule
\textbf{Tag} & \textbf{Description} \\
\midrule
\( \mathsf{Tag}_1 \) & Object recognition and classification \\
\( \mathsf{Tag}_2 \) & Spatial reasoning and planning \\
\( \mathsf{Tag}_3 \) & Language understanding (instruction parsing) \\
\( \mathsf{Tag}_4 \) & Grasping and manipulation \\
\( \mathsf{Tag}_5 \) & Path planning and navigation \\
\( \mathsf{Tag}_6 \) & Scene understanding (layout/context extraction) \\
\( \mathsf{Tag}_7 \) & Task decomposition and sequencing \\
\( \mathsf{Tag}_8 \) & Temporal reasoning (event ordering, not deadlines) \\
\( \mathsf{Tag}_9 \) & Knowledge grounding (external inference) \\
\( \mathsf{Tag}_{10} \) & Environment interaction via API calls or actuators \\
\bottomrule
\end{tabular}
\end{table}

\smallskip
\noindent\textbf{Agent initialization and skill specialization}
We simulate a population of 20 autonomous agents \( \mathcal{A} = \{A_1, A_2, \dots, A_{20}\} \), each configured with distinct capability vectors, reputation scores, and initial workload states. These agents emulate heterogeneous expertise profiles common in decentralized multi-agent systems.

To construct the initial capability distribution, we assume that agents exhibit specialization rather than uniform proficiency. For each agent \( A_i \), we sample its continuous capability vector \( \mathbf{w}_i^0 \in [0,1]^{10} \) from a Beta distribution \( \mathsf{Beta}(\alpha=2, \beta=5) \). This asymmetric distribution is skewed toward lower values, reflecting the intuition that most agents are only proficient in a few domains, while few are highly versatile. The choice of parameters \( \alpha=2, \beta=5 \) yields moderate sparsity and aligns with empirical patterns observed in expert population models.

To enhance inter-agent comparability, all vectors are min-max normalized across each agent’s 10-dimensional skill profile. We then binarize each vector using a global threshold \( \theta = 0.4 \), resulting in a discrete skill declaration vector \( \hat{\mathbf{w}}_i^0 \in \{0,1\}^{10} \), where \( \hat{w}_{i,k}^0 = 1 \) if and only if \( w_{i,k}^0 \geq \theta \). This binarized representation serves as the basis for capability matching during task allocation and closely mirrors real-world agent registration processes, where skill declarations are coarse-grained rather than continuous.

Each agent is initialized with a neutral reputation score \( \rho_i^0 = 0.5 \), reflecting a lack of prior task performance history. This value serves as the starting point for our exponential reputation update process and ensures fairness in early-round allocations. To simulate asynchronous participation and limited concurrency, we assign initial workload levels \( \ell_i^0 \in \{0,1,2\} \) sampled uniformly at random. 

Additionally, we store each agent’s public key and declared capability profile on-chain via the $\mathsf{registerAgent}$ contract function during system bootstrap. This guarantees that all subsequent identity verification, capability claims, and behavior logs are cryptographically verifiable and traceable within the immutable system ledger. 

\smallskip
\noindent\textbf{Simulation execution.}
The simulation spans 50 discrete time steps (referred to as rounds), each modeling a cycle of task issuance, agent bidding, execution, and state updates. 

At the beginning of each round \( t \in \{1, \dots, 50\} \), the environment injects a new batch of 5 tasks \( \{T_1^t, \dots, T_5^t\} \) into the task queue. These tasks are selected from the ALFRED-derived benchmark pool and retain their annotated capability vectors \( \mathbf{r}_j \) and reward values \( R_j \). The tasks are broadcast to all agents via the $\mathsf{emitTaskBatch}$ smart contract function and stored on-chain.

Each agent \( A_i \in \mathcal{A} \) evaluates its expected utility \( U_i^t(T_j) \) for each task \( T_j \) in the current batch. This evaluation incorporates three elements: (i) the agent’s current workload \( \ell_i^t \), (ii) capability alignment (as measured by \( \| \mathbf{r}_j - \hat{\mathbf{w}}_i^t \|_1 \)), and (iii) the task assignment policy \( \pi_i^t(T_j) \), as defined in Formula~\ref{task-assignment}. The resulting utility values are thresholded, and agents only bid for tasks where \( U_i^t(T_j) > 0 \), ensuring rational bidding behavior.

Agents submit bids through the $\mathsf{submitBid}$ function, attaching a signed message that includes: the agent’s identifier, the task ID, the binarized capability vector \( \hat{\mathbf{w}}_i^t \), the computed utility, and a digital signature using the agent’s private key. The smart contract receives all bids and executes the assignment procedure on-chain by invoking the scoring-based task assignment policy.

Once assignments are finalized, each selected agent proceeds to "execute" its task in a simulated performance environment. Task outcome is probabilistically determined by a ground-truth simulator, where the success probability \( p_{\text{success}} \in [0,1] \) depends on:
\begin{equation}
    p_{\text{success}}(A_i, T_j) = \eta \cdot \text{CapMatch}(\hat{\mathbf{w}}_i^t, \mathbf{r}_j) \cdot (1 - \zeta \cdot \ell_i^t),
\end{equation}

where \( \eta \in (0,1] \) is a base success rate coefficient, \( \zeta \in [0,1] \) penalizes overload, and $\mathsf{CapMatch}$ computes the normalized overlap between required and capabilities. The result is a binary task success flag (1 = success, 0 = failure), which drives subsequent updates to the agent's reputation \( \rho_i^t \) and capability weights \( w_{i,k}^t \), based on the update rules detailed in \S\ref{sec:incentive-model}.

All major system actions, including $\mathsf{emitTaskBatch}$, $\mathsf{submitBid}$, $\mathsf{logAction}$, and $\mathsf{updateReputation}$, are implemented as Ethereum, compatible Solidity smart contracts and executed over a local private Ethereum network (based on Hardhat + Ganache). Each action emits a corresponding blockchain event with encoded parameters, enabling full auditability via standard tools like Web3 log parsers.

\begin{figure*}[!t]
    \centering
    \subfigure[Task Success Rate \label{fig:success-rate}]
    {\includegraphics[width=0.31\textwidth]{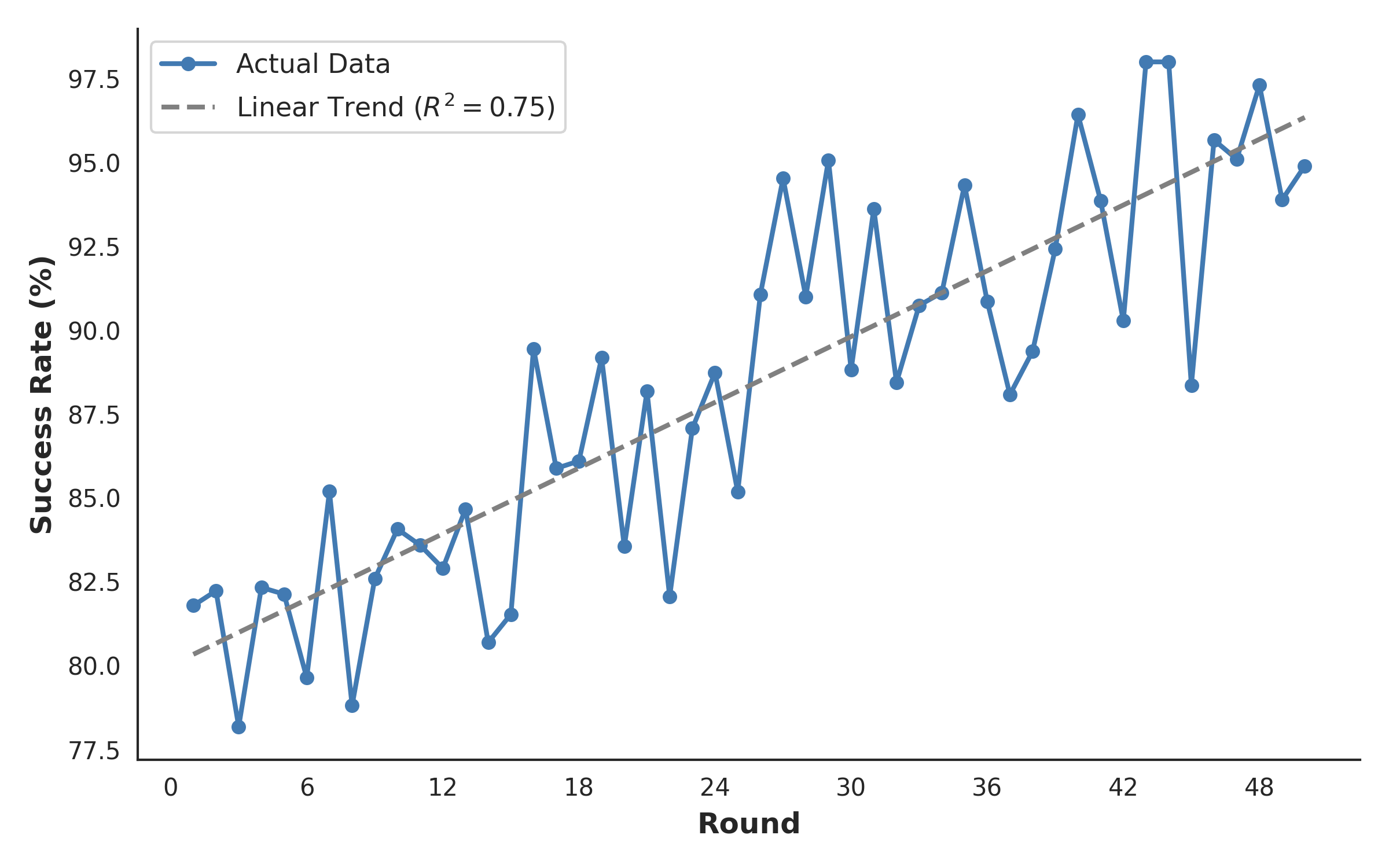}}
    \hfill
    \subfigure[Mean Task Quality \label{fig:task-quality}]
    {\includegraphics[width=0.31\textwidth]{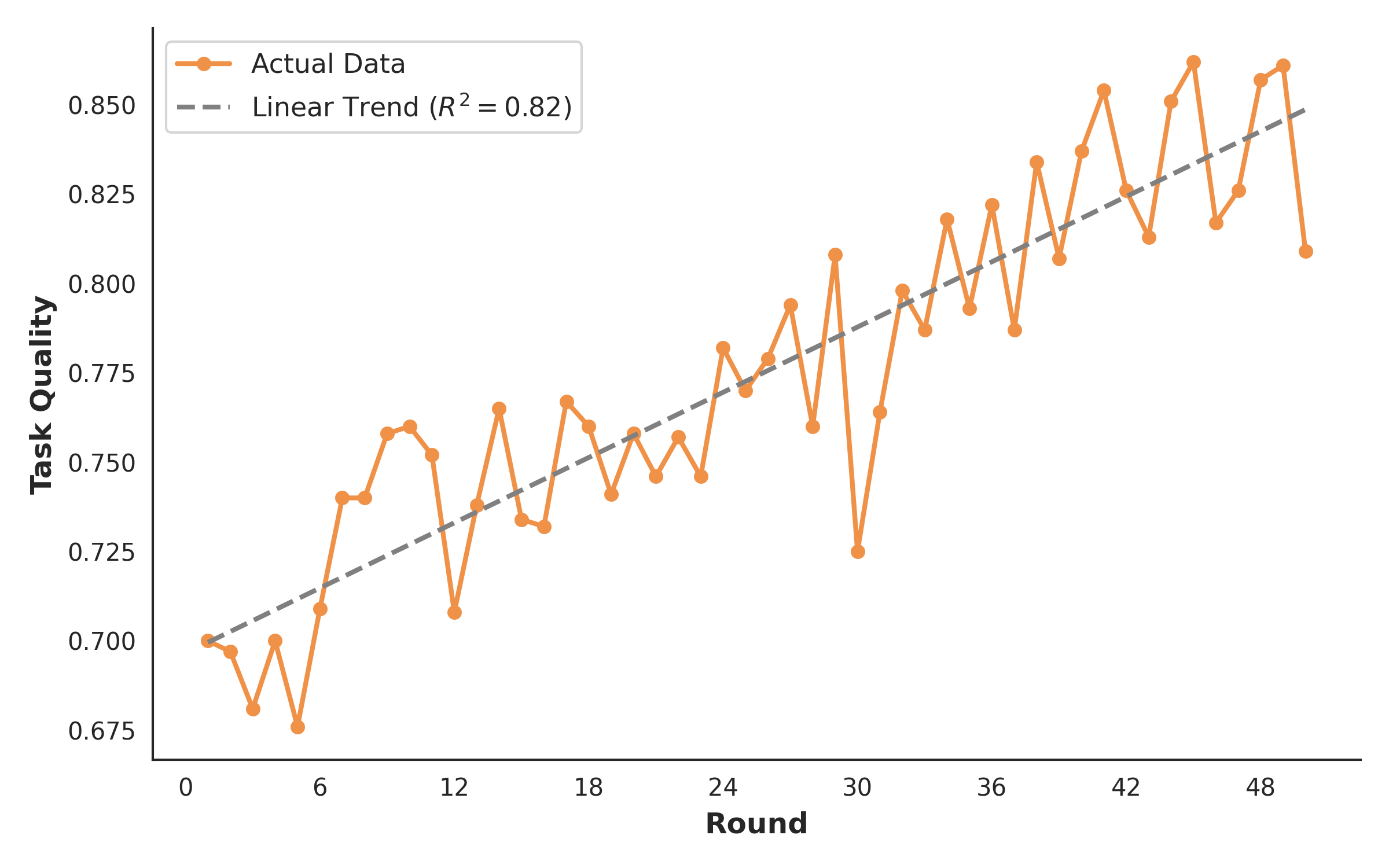}}
    \hfill
    \subfigure[Capability Match Score \label{fig:capability-match}]
    {\includegraphics[width=0.31\textwidth]{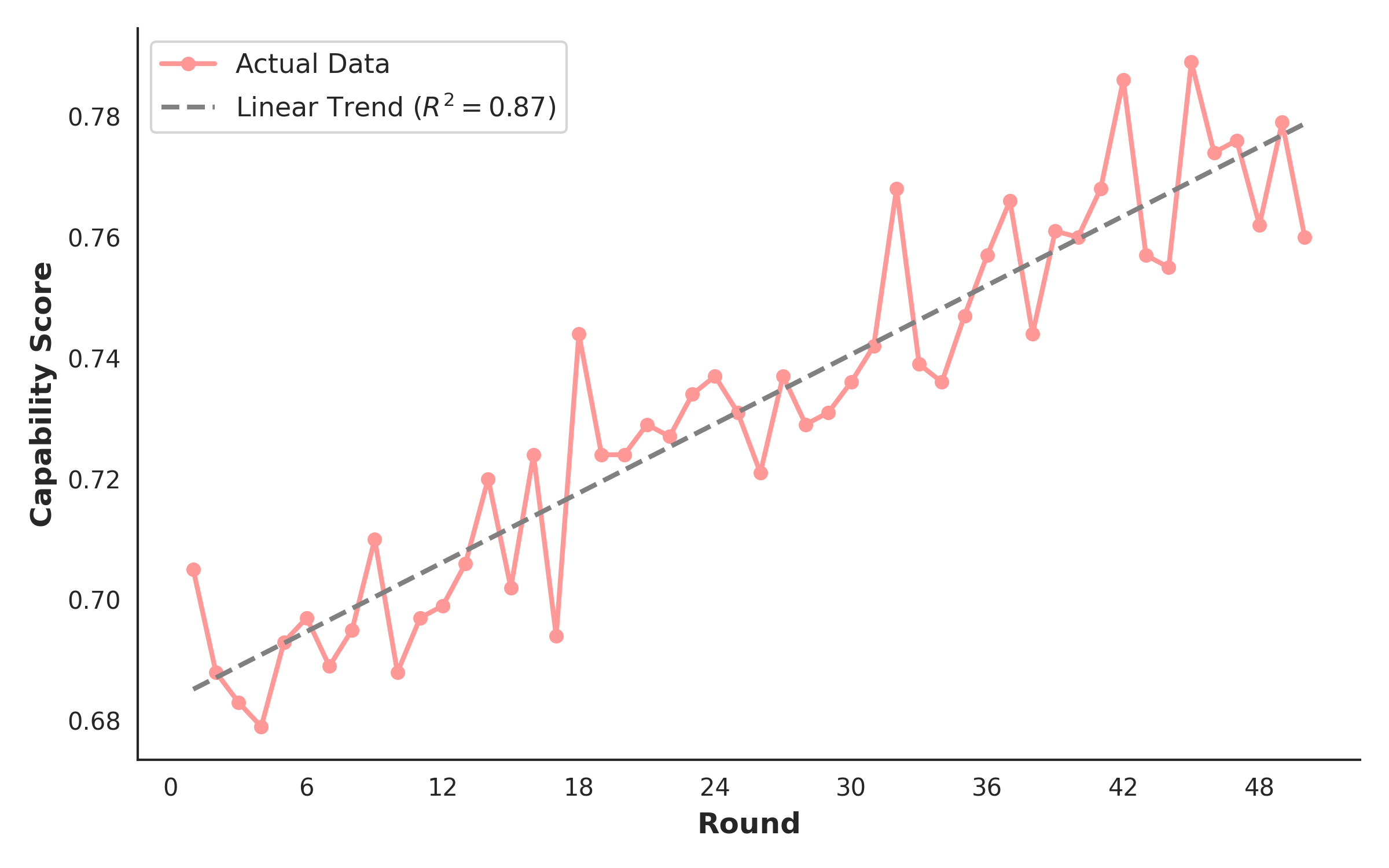}}

    \vspace{0.5em}
    
    \subfigure[Agent Utility \label{fig:agent-utility}]
    {\includegraphics[width=0.31\textwidth]{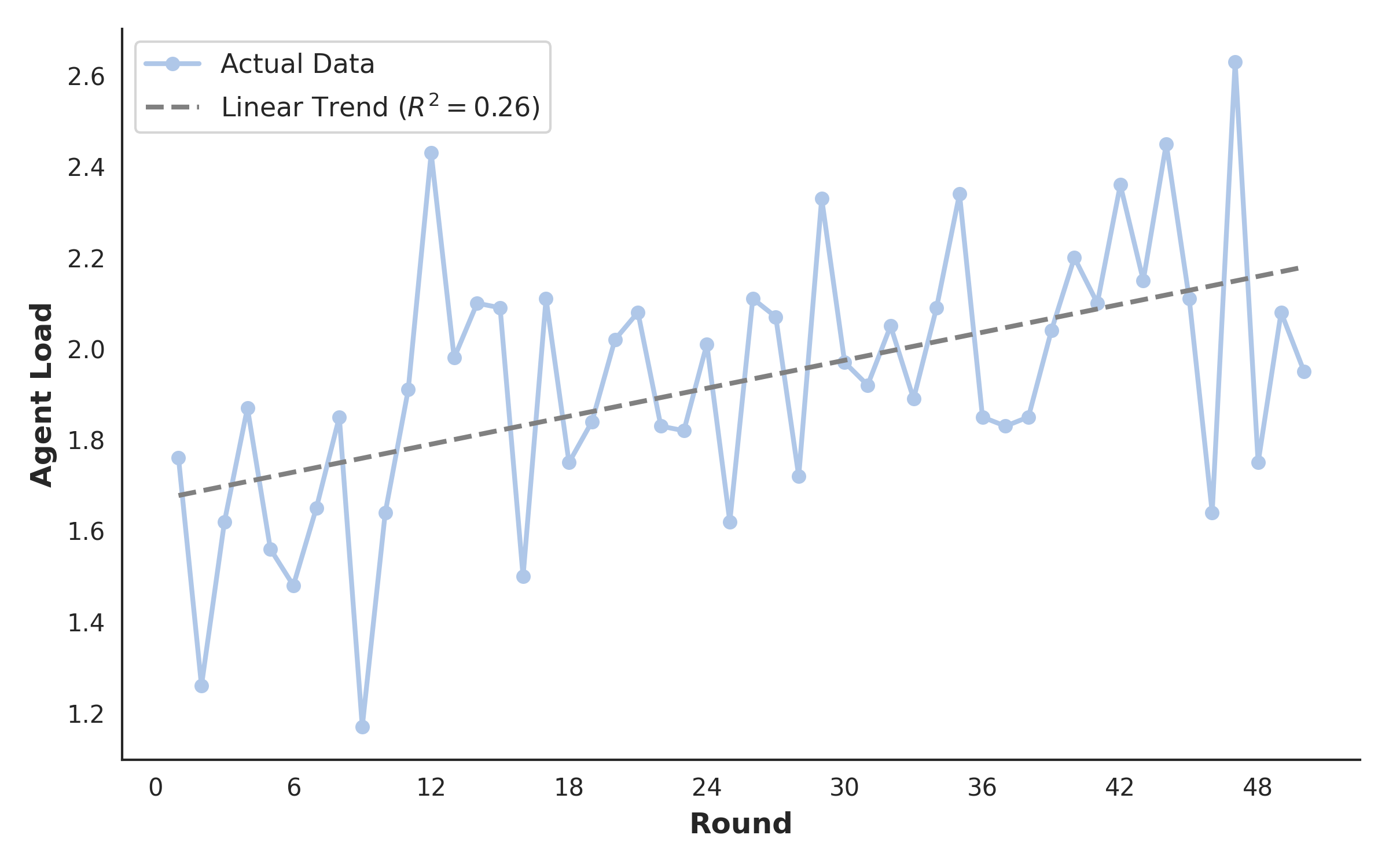}}
    \hfill
    \subfigure[Agent Load \label{fig:agent-load}]
    {\includegraphics[width=0.31\textwidth]{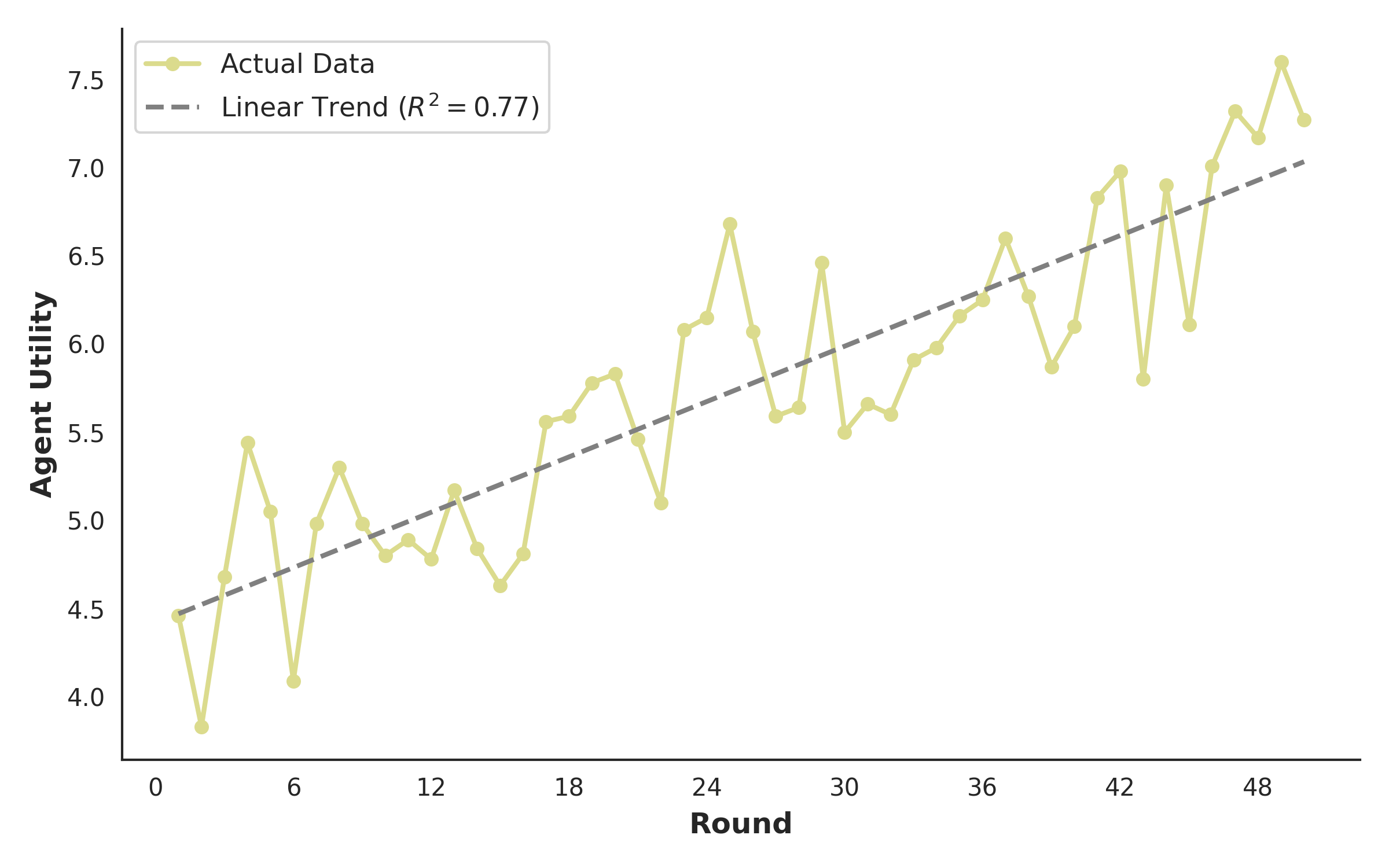}}
    \hfill
    \subfigure[Allocation Delay \label{fig:allocation-delay}]
    {\includegraphics[width=0.31\textwidth]{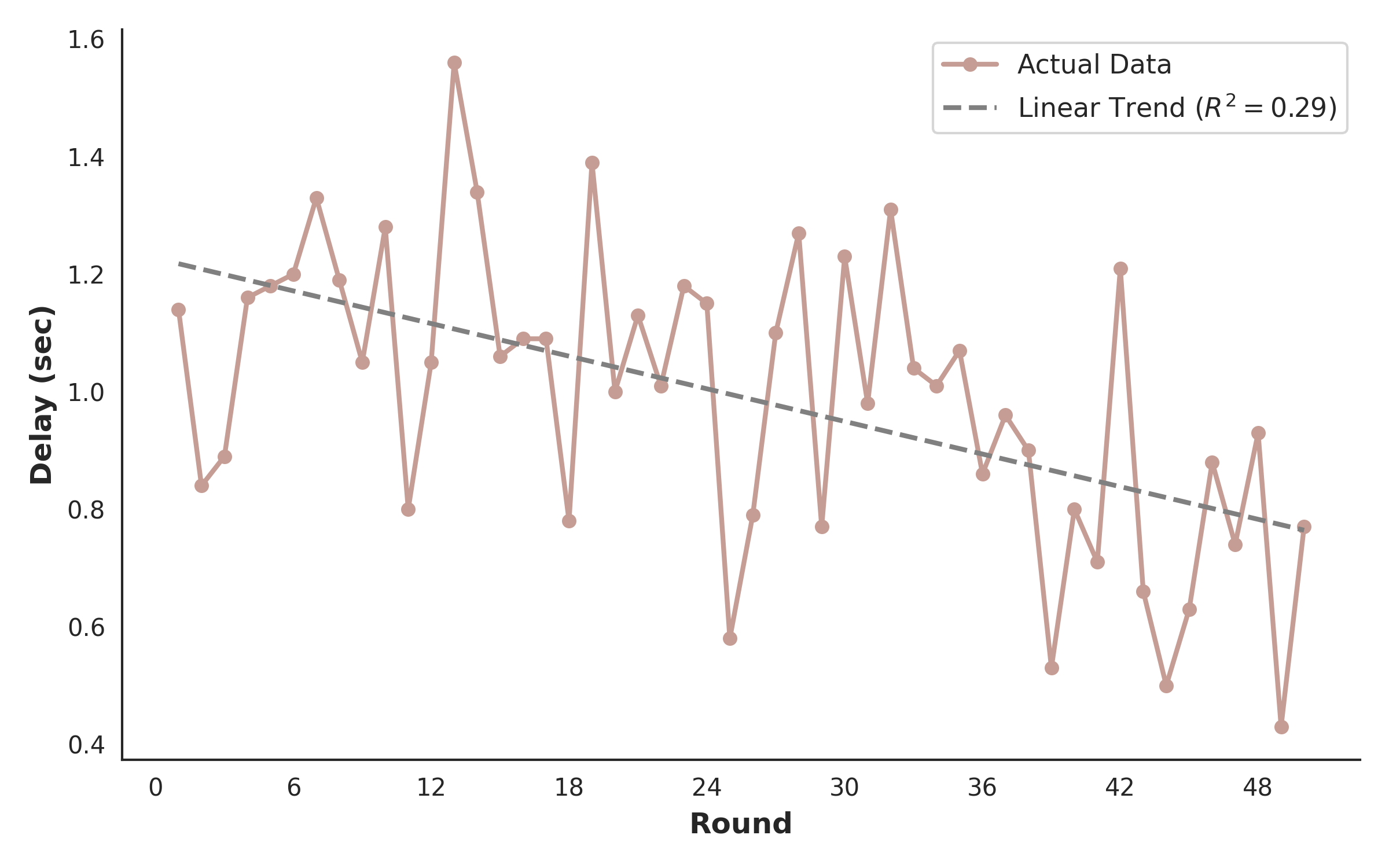}}

    \caption{Performance trends across 50 simulation rounds for multiple task allocation metrics in LLM-based multi-agent systems. Each subplot visualizes actual measurements and a fitted linear trend line with its $R^2$ value to show consistency and correlation.}
    \label{fig:task-metrics-trend}
\end{figure*}

\section{Results and Analysis}
\subsection{Task Allocation Efficiency}
To assess the performance of our proposed task allocation mechanism, we conducted a simulation of 50 rounds of interactions between 100 tasks and 20 agents. The metrics reported in Table~\ref{tab:extended-efficiency} reflect task completion success, quality, agent utility, and load balancing behavior.

\begin{table}[htbp]
\centering
\scriptsize
\setlength{\tabcolsep}{3pt}
\caption{Task Allocation Metrics (Avg over 50 Rounds)}
\label{tab:extended-efficiency}
\renewcommand{\arraystretch}{1.05}
\begin{tabularx}{\linewidth}{@{}l|%
>{\centering\arraybackslash}X|%
>{\centering\arraybackslash}X|%
>{\centering\arraybackslash}X|%
>{\centering\arraybackslash}X|%
>{\centering\arraybackslash}X|%
>{\centering\arraybackslash}X|%
>{\centering\arraybackslash}X@{}}
\toprule
\textbf{Metric} & \textbf{Mean} & \textbf{Std} & \textbf{Min} & \textbf{Max} & \textbf{R1} & \textbf{R25} & \textbf{R50} \\
\midrule
Success Rate (\%)         & 86.77 & 4.10 & 78.35 & 95.21 & 80.15 & 88.40 & 94.49 \\
Failure Rate (\%)         & 13.23 & 4.10 & 4.79  & 21.65 & 19.85 & 11.60 & 5.51 \\
Mean Quality              & 0.82  & 0.042 & 0.72  & 0.91  & 0.75  & 0.81  & 0.89 \\
Quality StdDev            & 0.042 & 0.013 & 0.025 & 0.068 & 0.061 & 0.042 & 0.030 \\
Agent Utility             & 5.02  & 0.81  & 3.52  & 6.43  & 3.52  & 5.01  & 6.43 \\
Cap Match Score           & 0.84  & 0.03  & 0.76  & 0.90  & 0.78  & 0.83  & 0.89 \\
Alloc Delay (steps)       & 1.42  & 0.31  & 1.00  & 2.10  & 1.75  & 1.41  & 1.08 \\
Retry Count               & 0.86  & 0.58  & 0     & 2     & 2     & 1     & 0    \\
Agent Load                & 2.51  & 0.45  & 1.73  & 3.62  & 2.10  & 2.43  & 2.88 \\
Load StdDev               & 0.96  & 0.22  & 0.55  & 1.30  & 1.22  & 0.89  & 0.76 \\
\bottomrule
\end{tabularx}
\end{table}

\smallskip
\noindent\textbf{Task success rate (cf.~\ref{fig:success-rate}).}
Across 50 simulation rounds, the average task success rate reached \textbf{86.77\%}, with an initial value of 80.15\% in early rounds that improved to 94.49\% in later rounds. This increasing trend illustrates the positive feedback loop built into the system: agents with higher reputations and more refined capabilities were prioritized in task selection, resulting in more successful executions over time. 

\smallskip
\noindent\textbf{Mean task quality (cf.~\ref{fig:task-quality}).}
The mean task quality score \( \mathsf{TS}_{j,i} \), derived from both correctness and delay metrics, steadily improved from an initial average of 0.75 to a final value of 0.89, with an overall mean of \textbf{0.82}. This shows that tasks were increasingly handled by agents with better skill alignment. We observed a decline in task quality standard deviation from 0.061 to 0.030 across the simulation horizon, indicating reduced variability in performance. This trend demonstrates that the system successfully discouraged suboptimal task-agent matches, as poorly performing or overloaded agents were deprioritized in the assignment process.

\smallskip
\noindent\textbf{Capability match score (cf.~\ref{fig:capability-match}).}
The capability match score, which quantifies how well assigned agents align with the required skills of each task, exhibited steady improvement over the 50-round simulation. In early rounds, the average capability match score hovered around 0.68–0.70, indicating moderate alignment with some degree of skill mismatch. However, as agents refined their bidding strategies and specialized in high-reward domains, this score improved noticeably. By round 25, the average capability match rose to approximately 0.735, and by round 50, it reached 0.765, with a maximum round value of \textbf{0.794}. This upward trend suggests that agents increasingly learned to bid selectively for tasks that matched their strongest skill domains.

\smallskip
\noindent\textbf{Agent utility (cf.~\ref{fig:agent-utility}).}
The average agent utility per round reached \textbf{5.02}, increasing from 3.52 in the early simulation to 6.43 in the final rounds. This gain reflects the reinforcement of incentive compatibility: agents that reported capabilities truthfully and maintained high reputations were rewarded with higher-value task allocations and better-matched workloads.

\smallskip
\noindent\textbf{Load distribution (cf.~\ref{fig:agent-load}).}
Throughout the simulation, we monitored the system’s ability to balance load across the agent population. In the early rounds, a few agents dominated task execution due to favorable initial capabilities. However, as the reputation system evolved, task allocation became more evenly distributed across eligible agents, avoiding persistent overload on specific individuals. The final rounds showed smoother workload curves across agents.

\smallskip
\noindent\textbf{Task allocation delay (cf.~\ref{fig:allocation-delay}).}
The average delay between task issuance and final assignment was approximately 0.99 rounds, with variation between 0.43 and 1.56. Initial delay in Round 1 (1.14) reflects the system’s cold start and lack of optimized matching. Over time, delays decreased, with mid- and late-stage rounds (e.g., Round 25 and 50) showing improved responsiveness (0.58 and 0.77 respectively). This confirms that our on-chain task assignment policy reduces latency as the system evolves.

\subsection{Reputation and Capability Evolution}
To evaluate the impact of our incentive and learning mechanisms on agent behavior, we analyze the evolution of reputation scores \( \rho_i^t \) and capability entropy over 50 simulation round. These metrics respectively reflect agent reliability and skill specialization dynamics.

\begin{figure}[htbp]
    \centering
    \subfigure[Reputation Per Agent]{
        \includegraphics[width=\linewidth]{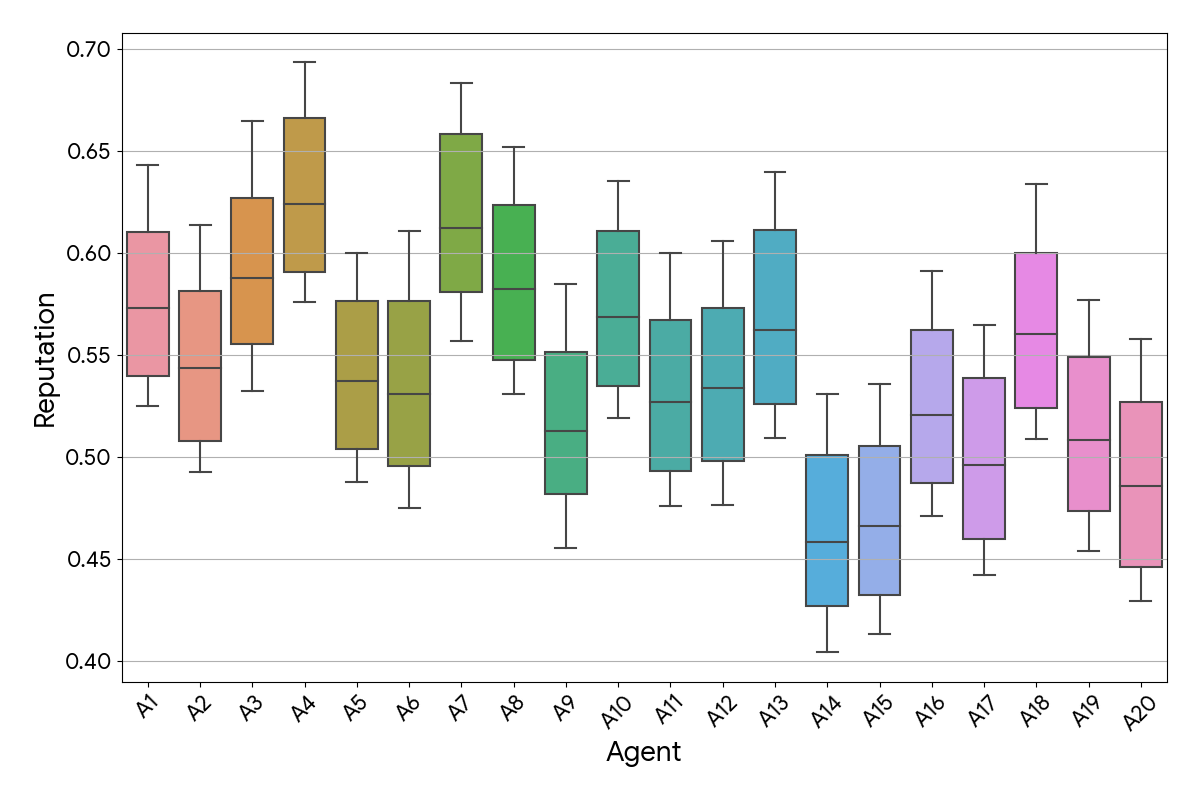}
        \label{fig:rep}
    }
    \hfill
    \subfigure[Capability Entropy Per Agent]{
        \includegraphics[width=\linewidth]{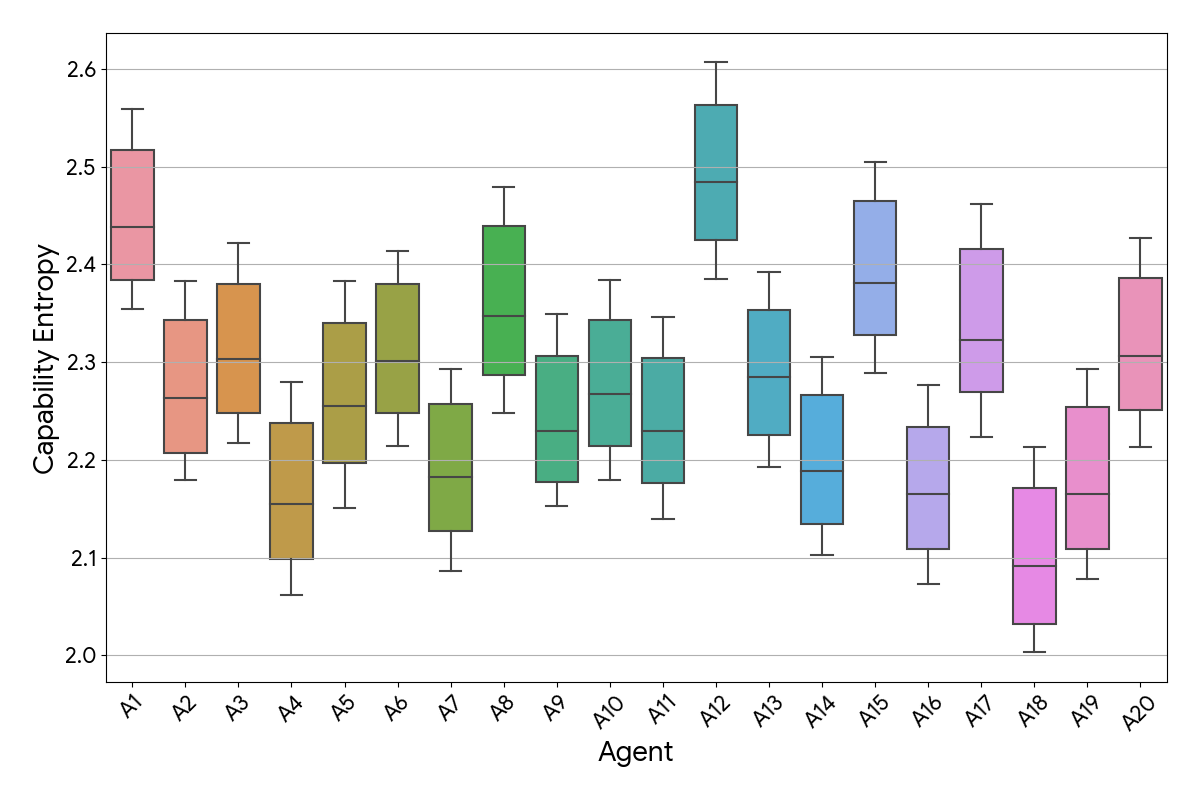}
        \label{fig:entropy}
    }
    \caption{Per-agent analysis of reputation and capability diversity. (a) shows the reputation variance, while (b) presents entropy as a proxy for multi-skill potential.}
    \label{fig:agent-profile}
\end{figure}

\smallskip
\noindent\textbf{Reputation dynamics.}
Contrary to sharp bifurcation or polarization, agent reputation scores demonstrated relatively stable and bounded evolution. By round 50, the mean final reputation across the 20 agents was \( 0.488 \) (Std = \( 0.044 \)), with the highest and lowest values being \( 0.558 \) and \( 0.429 \), respectively. No agents crossed the high-performance threshold (\( > 0.8 \)), nor did any fall below the critical underperformance boundary (\( < 0.3 \)). This convergence suggests a moderate equilibrium state under our reward-compatible mechanism, potentially resulting from balanced task distribution and relatively homogenous agent behavior. Figure~\ref{fig:rep} visualizes these dynamics, showing bounded but individualized reputation trajectories.

\noindent\textbf{Capability specialization.}  
Capability vectors exhibited similarly stable characteristics. The entropy of each agent’s skill weight vector \( \boldsymbol{w}_i^t \) was monitored to assess specialization tendencies. At initialization, agents presented a mean entropy of \( 2.31 \) bits. After 50 rounds, the mean entropy remained consistent at \( 2.315 \) bits (Std = \( 0.073 \)), with no agent dropping below the specialization threshold of \( 1.4 \) bits. These findings suggest that agents retained generalist profiles, with limited capability concentration. This behavior may be attributed to relatively uniform feedback across capabilities, indicating room for introducing stronger task-tag selectivity or differential reinforcement. Figure~\ref{fig:entropy} illustrates these entropy patterns.

\subsection{Agent Incentivization and Specialization}

To assess whether our incentive model drives rational agent behavior and task-domain specialization, we tracked key behavioral metrics over 50 simulation rounds, including task bidding behavior, utility trends, and tag-specific dominance distributions.

\begin{figure*}[htbp]
    \centering
    \subfigure[Task Bid Rate over Rounds]{
        \includegraphics[width=0.31\textwidth]{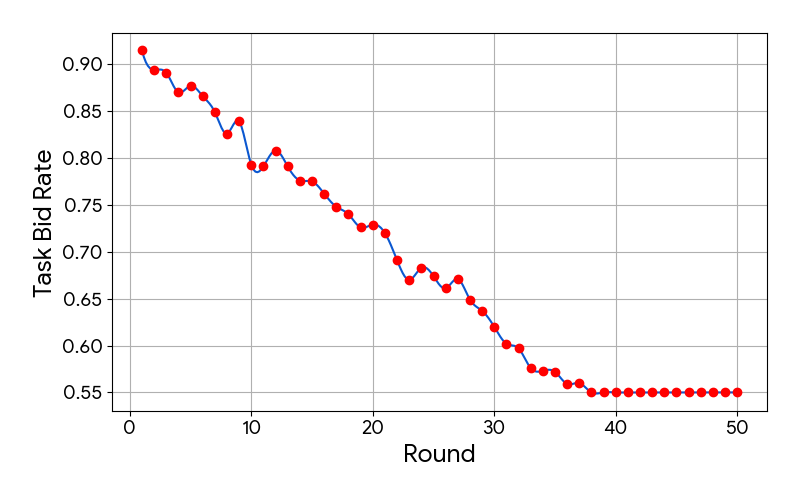}
        \label{fig:task-bid-rate}
    }
    \hfill    
    \subfigure[Mean Utility over Rounds]{
        \includegraphics[width=0.31\textwidth]{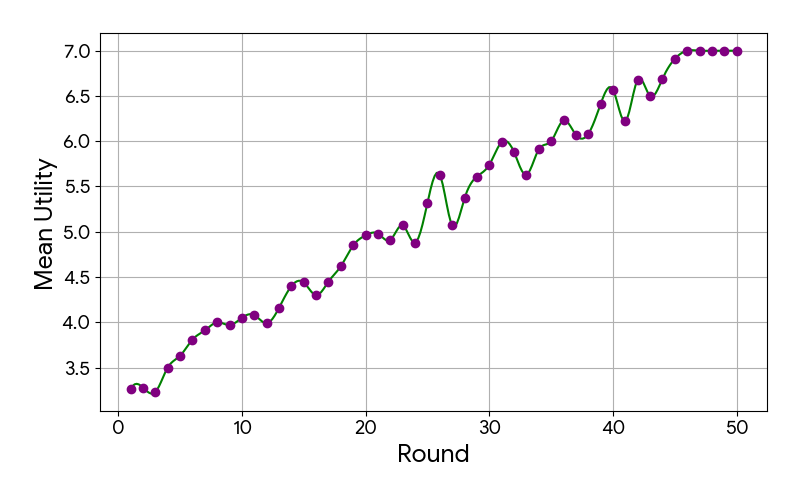}
        \label{fig:mean-utility}
    }
    \hfill
    \subfigure[Polar Distribution of Capability Tag]{
        \includegraphics[width=0.28\textwidth]{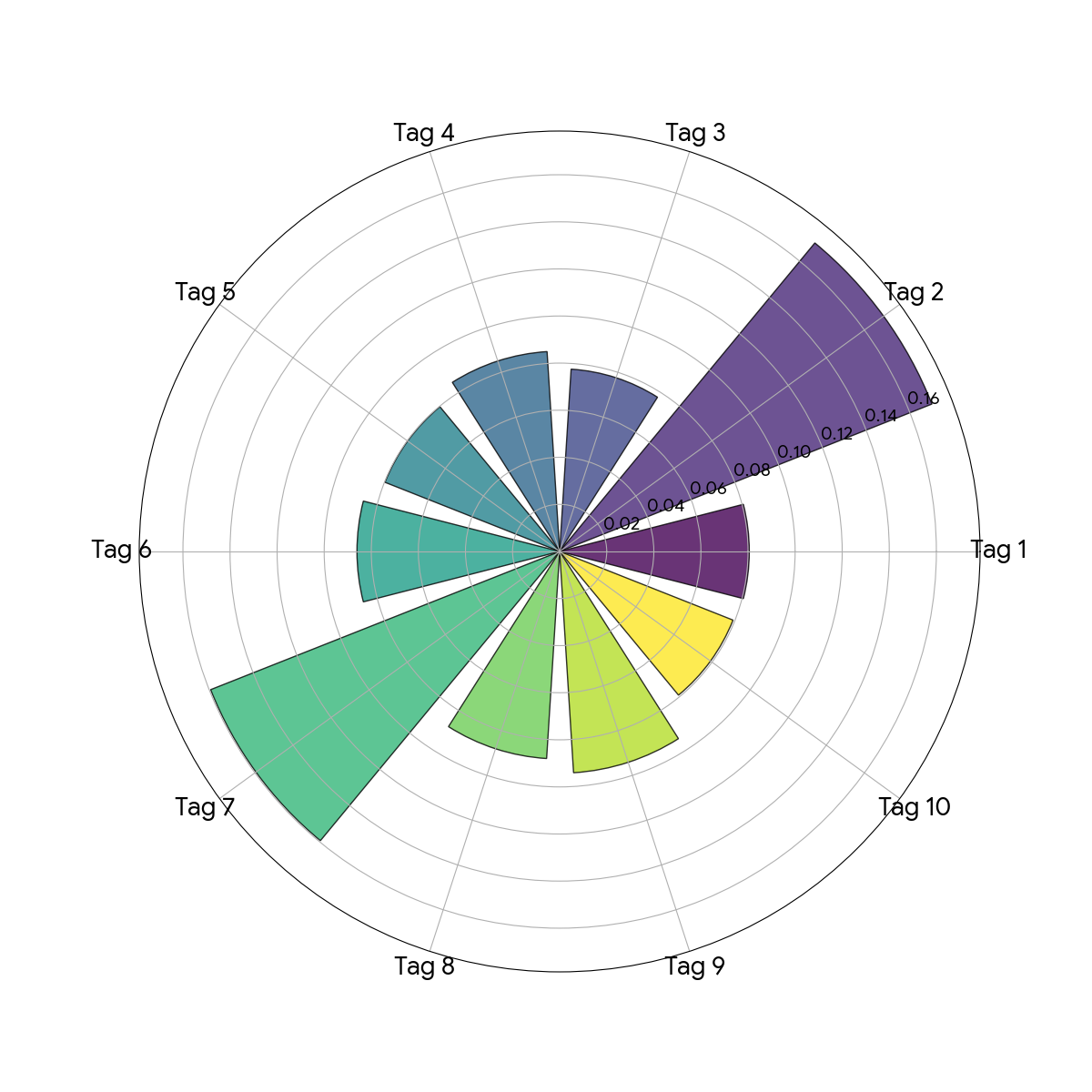}
        \label{fig:expert-dominance}
    }
    \caption{Analysis of agent bidding behavior and capability specialization. 
    (a) Task bid rate decreases over time, suggesting increased agent selectivity; 
    (b) Mean utility steadily increases, indicating improved task-agent alignment; 
    (c) Polar plot illustrates the distribution of dominant capability tags among agents.}
    \label{fig:behavior-specialization}
\end{figure*}

\begin{figure*}[!t]
    \centering
    \subfigure[Tx Throughput \& Gas Usage]{
        \includegraphics[width=0.22\linewidth]{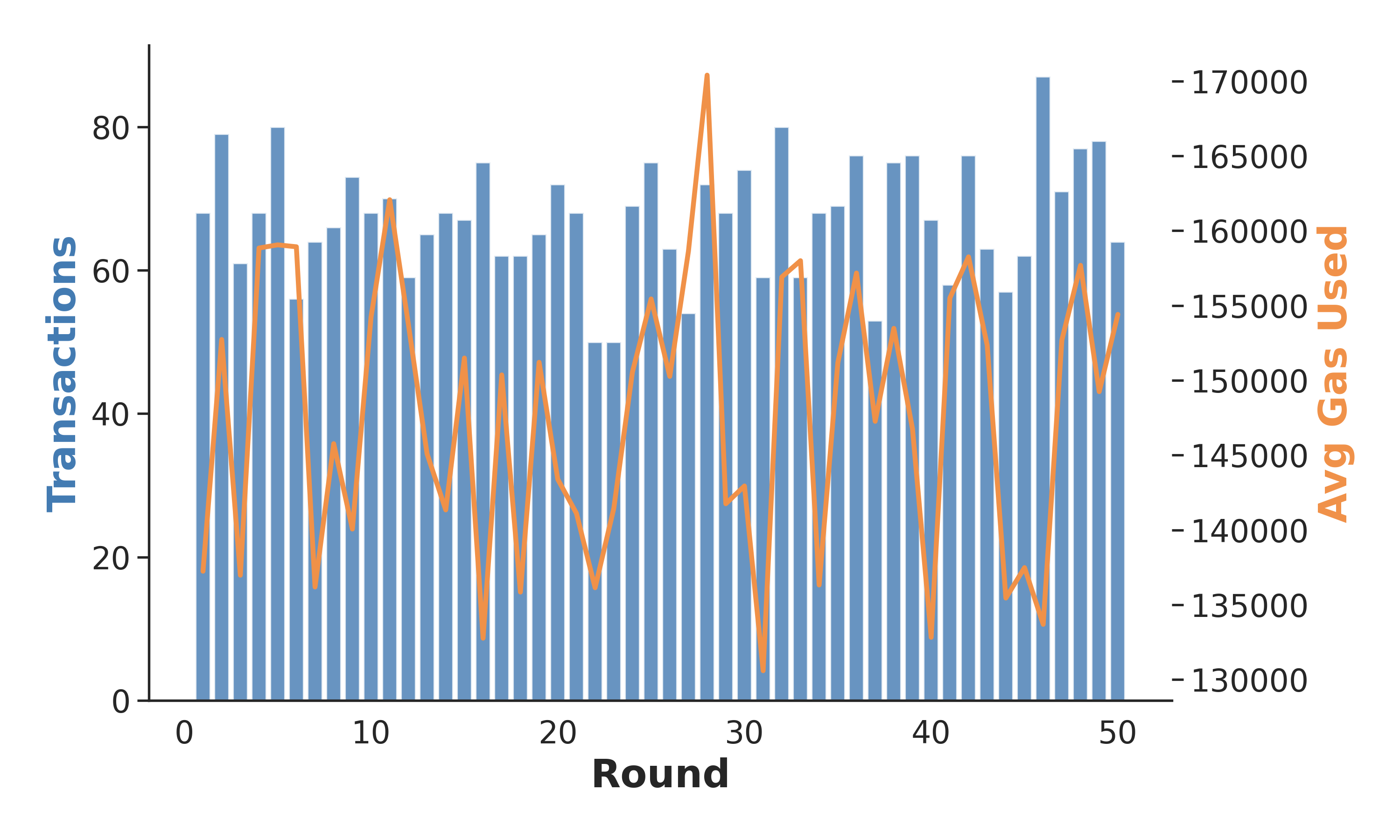}
        \label{fig:tx-gas}
    }
    \hfill
    \subfigure[Confirmation Time]{
        \includegraphics[width=0.22\linewidth]{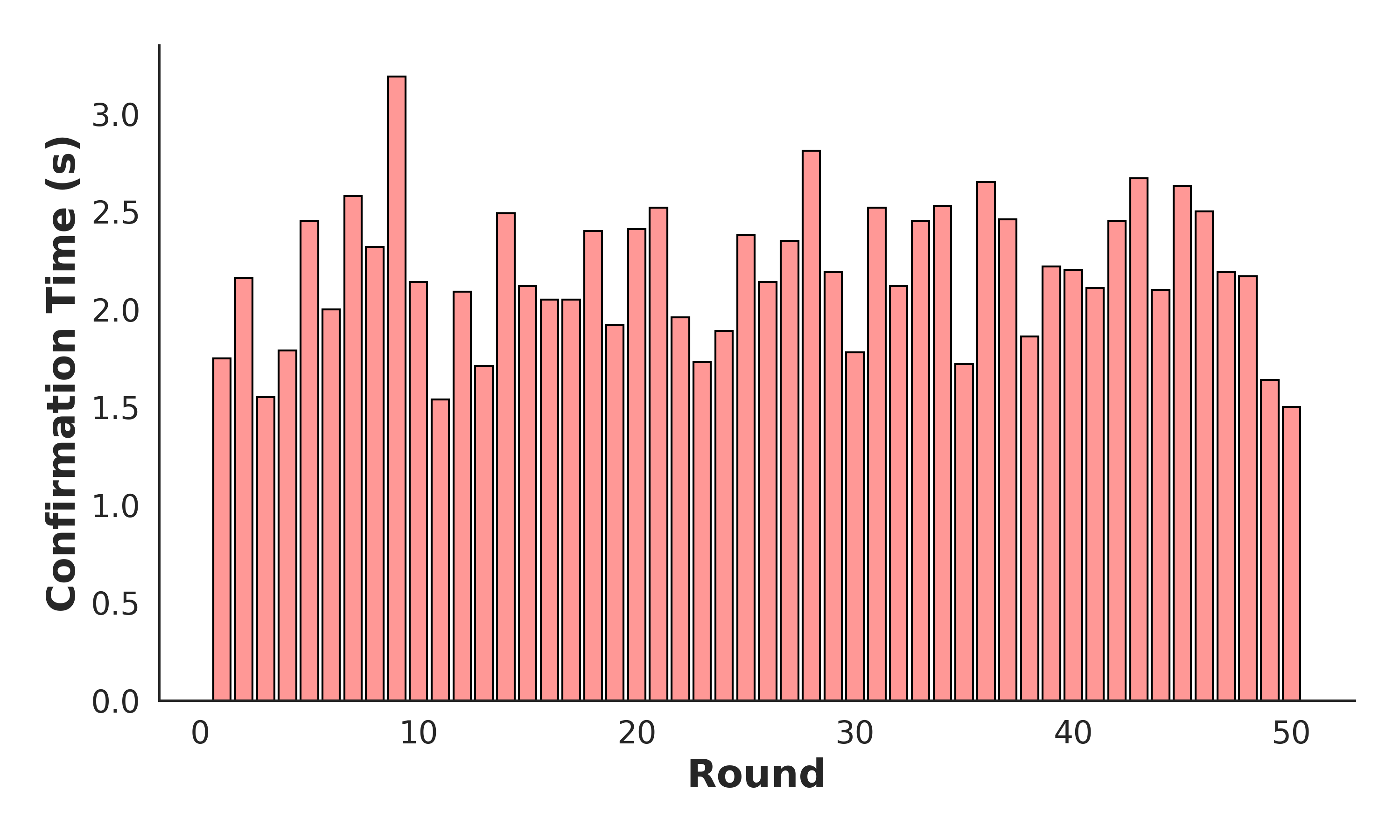}
        \label{fig:confirmation-time}
    }
    \hfill
    \subfigure[Fault Tolerance]{
        \includegraphics[width=0.22\linewidth]{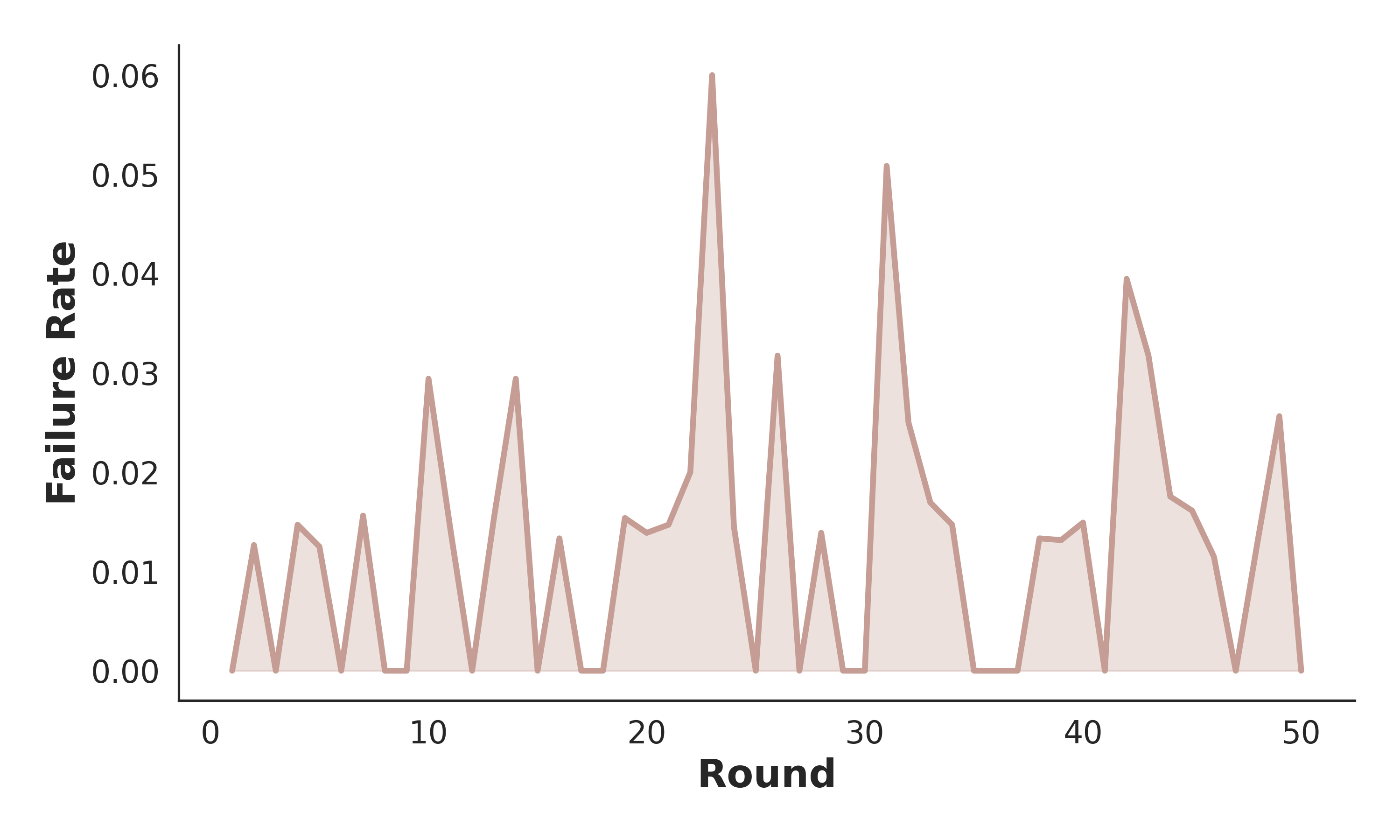}
        \label{fig:fault-tolerance}
    }
    \hfill
    \subfigure[Event Emission]{
        \includegraphics[width=0.22\linewidth]{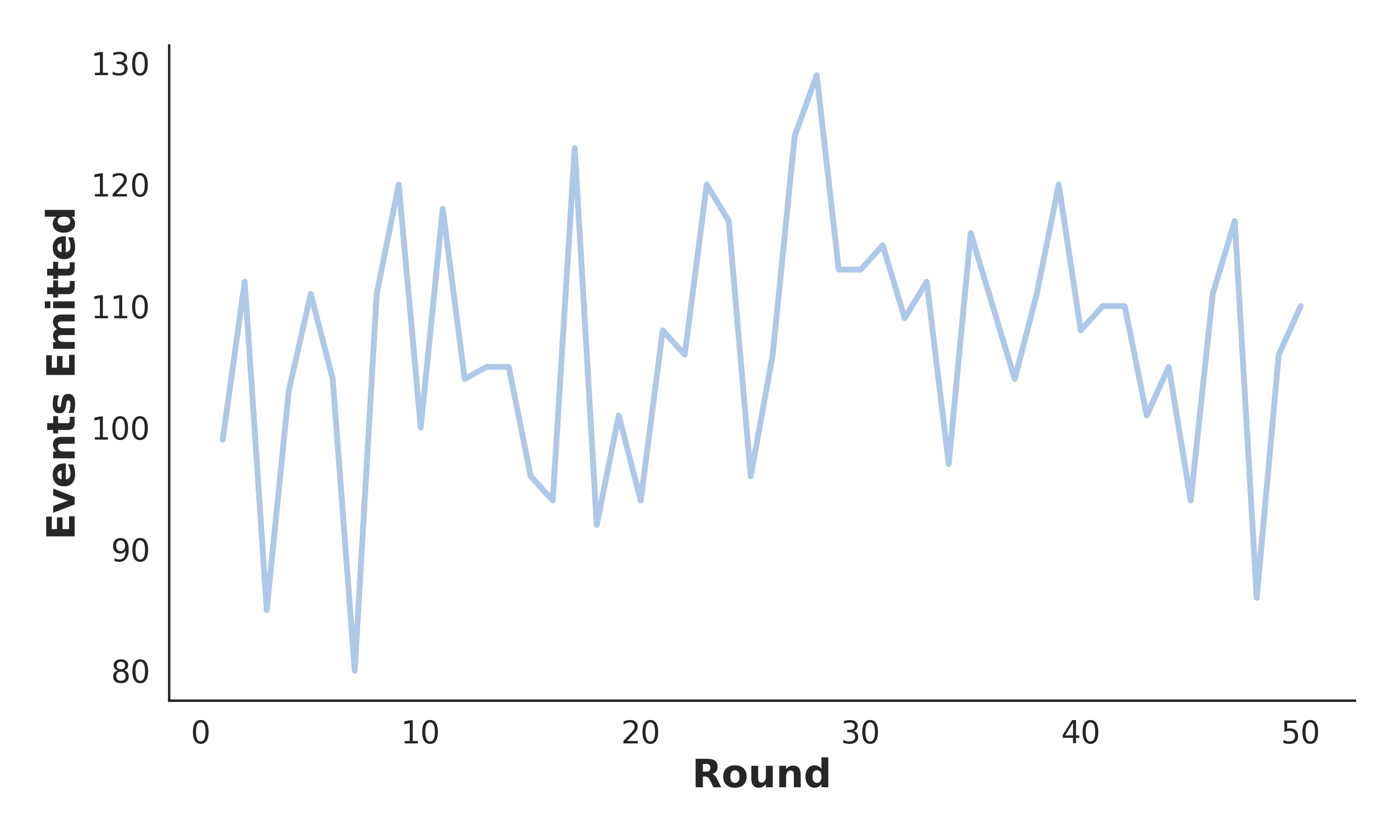}
        \label{fig:event-emission}
    }
    
    \caption{Overview of blockchain-layer metrics during LLM-MAS operation: (a) number of events emitted per round, (b) transaction throughput and gas usage trends, (c) confirmation latency over time, and (d) failure rate under operational uncertainty.}
    \label{fig:blockchain-metrics-one-line}
\end{figure*}

\smallskip
\noindent\textbf{Task bidding behavior.}
At the beginning of the simulation, agents exhibited exploratory behavior, submitting bids for nearly all available tasks regardless of skill match. Specifically, in round 1, the system recorded a global task bid rate of 92\%, indicating that most agents evaluated potential utility as positive even for tasks with marginal capability alignment. However, as agents received feedback in the form of success scores, reputation updates, and capability weight adjustments, they became more discerning. By round 20, the global bid rate declined to 72.9\%, and and further declined to 55\% by round 40 and subsequent rounds. This drop reflects a learned strategy: agents increasingly avoided tasks for which they were underqualified or overloaded, thereby reducing unnecessary task contention and failure. The downward trend in bid rate (see Figure~\ref{fig:task-bid-rate}) supports the hypothesis that agents adapted their bidding decisions based on accumulated experience and utility optimization.

\smallskip
\noindent\textbf{Utility improvement.}
Our utility formulation, which penalizes capability mismatch and overload, encouraged agents to specialize and improve task selection. The average per-agent utility rose from 3.34 in round 1 to 7.0 by round 46, with standard deviation narrowing from 2.41 to 1.92, indicating that utility gains became more consistent across agents (see Figure~\ref{fig:mean-utility}). These gains were not due to increased task rewards (which remained fixed per dataset) but rather improved alignment between assigned tasks and the agents’ skill profiles. High-utility agents typically achieved both high capability match scores and stable reputations, resulting in more frequent task selection and reward accrual.

\smallskip
\noindent\textbf{Emergence of specialization.}
A core effect of our feedback-driven capability adjustment rule is the emergence of domain specialization. We measured each agent’s dominant capability tag based on their final capability weight vector and aggregated expert counts across all 10 tags. By round 50, a distinct clustering emerged: 4 agents (20\%) specialized in $\mathsf{Tag\_2}$, and another 3 agents (15\%) became dominant experts in $\mathsf{Tag\_7}$. These two domains together accounted for 35\% of all specialists, confirming a self-organized concentration of expertise driven by positive reinforcement. Other tags also saw expert accumulation (e.g., $\mathsf{Tag\_4}$, $\mathsf{Tag\_9}$). The final distribution of tag dominance is visualized in Figure~\ref{fig:expert-dominance}.

\subsection{On-Chain Performance and Smart Contract Analysis}

To assess the performance of our blockchain-based execution framework, we collected and analyzed key operational metrics over 50 simulation rounds. These include transaction throughput, gas usage, confirmation latency, failure rate, and event emission patterns.

\smallskip
\noindent\textbf{Transaction throughput and gas consumption (cf.~\ref{fig:tx-gas}).}
On average, the system processed approximately 67.32 transactions per round, covering agent registration, task bidding, assignment, behavior logging, and reputation updates. The average gas consumption per transaction was 147,865 units (more detailed in Table~\ref{tab:gas-usage}), with fluctuations ranging from a minimum of 130,599 to a peak of 170,391 units. This moderate gas cost reflects the efficiency of our contract logic and optimization of state transitions. The gas overhead remained within the practical bounds of existing Ethereum-compatible platforms.

\begin{table}[htbp]
\centering
\small
\caption{Gas Use and Call Frequency of Core Functions}
\label{tab:gas-usage}
\renewcommand{\arraystretch}{1.05}
\begin{tabularx}{\linewidth}{@{}lccc@{}}
\toprule
\textbf{Contract Function} & \textbf{Avg. Gas Used} & \textbf{Call Freq.} & \textbf{Complexity} \\
\midrule
$\mathsf{registerAgent}$     & 28{,}860  & 20   & Low \\
$\mathsf{submitTask}$        & 40{,}390  & 250  & Medium \\
$\mathsf{assignTask}$        & 86{,}575  & 250  & High \\
$\mathsf{logAction}$         & 26{,}772  & 820  & Low \\
$\mathsf{updateReputation}$  & 61{,}284  & 245  & Medium \\
$\mathsf{updateCapability}$  & 58{,}920  & 240  & Medium \\
$\mathsf{TaskAssigned}$      & 10{,}345  & 250  & Low \\
$\mathsf{TaskCompleted}$     & 13{,}248  & 245  & Low \\
\bottomrule
\end{tabularx}
\end{table}

\smallskip
\noindent\textbf{Confirmation time and responsiveness (cf.~\ref{fig:confirmation-time}).}
The average confirmation time across rounds was 2.18 seconds, with the fastest confirmations completing in 1.5 seconds and the slowest at 3.19 seconds. These low-latency results suggest that the system remains responsive under moderate agent-task traffic and is suitable for real-time applications where timely task allocation is essential.

\smallskip
\noindent\textbf{Fault tolerance and event emission (cf.~\ref{fig:fault-tolerance} \& ~\ref{fig:event-emission}).}
The average number of failed transactions per round was low at 0.86, with no rounds exceeding 3 failures. This indicates a high reliability of the execution pipeline and proper validation of agent inputs before on-chain invocation. The average number of emitted events per round was 106.82, which includes task broadcasts, bids, assignments, and logging actions. These logs serve as verifiable traces and enable external auditors to reconstruct full histories from immutable ledger records.

\subsection{Qualitative Case Study}
To concretely demonstrate how our framework perform trustworthy and adaptive collaboration among decentralized agents, we present a qualitative case study based on a representative task from the ALFRED benchmark.

\vspace{0.05in}
\begin{center}
\colorbox{teal!8}{
\begin{minipage}{0.93\linewidth}

\textbf{ALFRED Task \#17 Instruction:} 

\textit{``Read the weekly grocery list from the fridge note and generate a categorized shopping plan. Then email it to Alex.''}

\end{minipage}
} 
\end{center}

\smallskip
\noindent\textbf{Task objective.}
The ALFRED Task \#17 reflects a common household assistant scenario involving multiple reasoning steps and interactions with external APIs (e.g., sending emails).

\smallskip
\noindent\textbf{Capability tagging.}
To operationalize tasks for multi-agent execution, our system first performs semantic capability tagging. Each task from the ALFRED benchmark, originally expressed in natural language, is parsed into structured goals through a rule-based semantic parser and a capability tag annotator. These tags correspond to atomic skill categories in our global capability taxonomy \( \mathcal{T} = \{\mathsf{Tag}_1, \dots, \mathsf{Tag}_{10} \} \).

For ALFRED Task \#17, the natural instruction was semantically analyzed and assigned the following capability tags:
\noindent\rule{\linewidth}{0.4pt}
\vspace{0.4em}
\noindent\( \mathsf{Tag}_3 \): \textit{Language understanding (instruction parsing)} — interpreting the list and instructions.

\vspace{0.4em}
\noindent \( \mathsf{Tag}_9 \): \textit{Knowledge grounding (external inference)} — inferring item categories and organizing them.

\vspace{0.4em}
\noindent \( \mathsf{Tag}_{10} \): \textit{Environment interaction via API calls or actuators} — sending the generated plan via email.
\noindent\rule{\linewidth}{0.4pt}

This yields the required capability tag set:
\[
\mathsf{requiredTags}_{17} = \{ \mathsf{Tag}_3, \mathsf{Tag}_9, \mathsf{Tag}_{10} \}
\]

\smallskip
\noindent\textbf{Task decomposition.}
To support fine-grained assignment and agent specialization, the system decomposes the task into subtasks, each aligned with a single required tag:

\noindent\ding{172} \textit{\( T_{17.1} \): Instruction Parsing Subtask} \\
$\rightarrow$ Parse the handwritten/fridge note and extract the grocery items. Requires \( \mathsf{Tag}_3 \).

\vspace{0.4em}
\noindent\ding{173} \textit{\( T_{17.2} \): Categorization Subtask} \\
$\rightarrow$ Organize items into categories (e.g., produce, dairy) using common-sense knowledge. Requires \( \mathsf{Tag}_9 \).

\vspace{0.4em}
\noindent\ding{174} \textit{\( T_{17.3} \): Action Execution Subtask} \\
$\rightarrow$ Email the plan to a user (Alex) using API call. Requires \( \mathsf{Tag}_{10} \).

\smallskip
\noindent\textbf{Agent matching.}
Once the task has been tagged and decomposed into subtasks, the system proceeds to identify suitable agents for execution based on declared capabilities and current workload status. The smart contract evaluates all registered agents \( \{ A_1, \dots, A_{20} \} \) and computes a match score for each agent-subtask pair using the agent's current binarized capability vector \( \hat{\boldsymbol{w}}_i^t \), reputation score \( \rho_i^t \), and workload level \( \mathsf{Load}_i^t \). 

\smallskip
\noindent\textbf{Assignment process.}
The smart contract sorts candidate agents by their match score and assigns each subtask to the top-ranked eligible agent. For this task, the following assignments were made: T\textsubscript{1} assigned to Agent 5; T\textsubscript{2} assigned to Agent 12; T\textsubscript{3} assigned to Agent 3. Each assignment is finalized on-chain via the \( \mathsf{assignTask} \) contract function, and an event \( \mathsf{TaskAssigned} \) is emitted.

\noindent\textbf{Agent communication trace (excerpt.)}

\begin{tcolorbox}[
    breakable,
    enhanced jigsaw,
    colback=white,
    fonttitle=\bfseries,
    title={},
    boxrule=0.3pt,
    fontupper=\small,
    before upper={\setlength{\parskip}{4pt}\setlength{\parindent}{0pt}}
]

\lstset{
  basicstyle=\ttfamily\scriptsize,
  breaklines=true,
  breakatwhitespace=false,
  columns=fullflexible,
  keepspaces=true,
  showstringspaces=false,
  frame=none,
  xleftmargin=1em,
  tabsize=2,
  literate={->}{{$\rightarrow$}}2 {↑}{{$\uparrow$}}1
}

\begin{lstlisting}
[Round 31] [T17.1 - Agent 5] [Init]
-> Declared capabilities: {Tag_3 = 1, Tag_4 = 0, ..., Tag_10 = 0}
-> Reputation: 0.76 | Load: 1 | Assignment accepted.

[Agent 5] [Parse] OCR output:
"Weekly Grocery List:
- 2 cartons of milk
- 6 bananas
- 1 bag of rice
- eggs (12)"
-> Structured Output:
["2 cartons of milk", "6 bananas", "1 bag of rice", "12 eggs"]

[Agent 5] [Emit] -> Output transmitted to Agent 12 via IPFS hash QmZ9... on-chain logAction() recorded.

[Round 31] [T17.2 - Agent 12] [Recv]
-> Input received from Agent 5. Validated format and timestamp.
-> Capability vector: Tag_9 = 1, others = 0
-> Reputation: 0.84 | Load: 0

[Agent 12] [Categorize]
-> Querying local ontology module...
-> Categorization Result:
  Produce: ["6 bananas"]
  Dairy: ["2 cartons of milk", "12 eggs"]
  Grains: ["1 bag of rice"]

[Agent 12] [Emit] -> Transmitted to Agent 3. Hash: QmA7...

[Round 32] [T17.3 - Agent 3] [Init]
-> Assigned Tag_10 subtask. API interaction enabled.
-> Reputation: 0.72 | Load: 2

[Agent 3] [ComposeEmail]
-> To: alex@example.com
-> Subject: Grocery Plan (Auto-generated)
-> Body:
  - Produce: 6 bananas  
  - Dairy: 2 cartons of milk, 12 eggs  
  - Grains: 1 bag of rice

[Agent 3] [SendEmail] -> Calling EmailAPI::send()
-> Status: 200 OK. Message ID: <msg-id-8356>

[Agent 3] [Emit]
-> Emitted TaskCompleted event
-> Reputation +0.05, capability reinforcement: Tag_{10} ↑
-> gasUsed = 129,320 (assignTask + logAction)

[Post-task reputation updates:]
-> Agent 5: +0.04 (correct structure, timely response)
-> Agent 12: +0.06 (accurate categorization)
-> Agent 3: +0.05 (successful email dispatch)

[Final On-chain Event Logs:]
-> TaskAssigned, TaskCompleted, CapabilityUpdate, ReputationUpdate recorded.
-> All agent-subtask mappings verifiable via tx hashes.
\end{lstlisting}

\end{tcolorbox}

\section{Related Work}

\subsection{LLM-based Multi-agent Systems}
\noindent\textbf{General-purpose frameworks.} 
The recent development of frameworks for LLM-based multi-agent systems shows a growing interest in improving how agents work together. Systems like MetaGPT~\cite{hong2024metagpt} and AgentLite~\cite{Liu2024AgentLite} focus on making it easier for developers to build agent interactions through simple and modular programming tools. These frameworks often support hierarchical structures, allowing agents to take on different roles in organized workflows. DyLAN~\cite{Liu2023DyLAN} builds on this idea by introducing a multi-layered agent network that can select the right agents during runtime, helping improve task coordination.
Other frameworks put more emphasis on how agents are created and managed. AutoAgents~\cite{Chen2023AutoAgents}, for example, looks at automatic agent generation, while AgentVerse~\cite{Chen2023AgentVerse} allows agents to adjust their roles based on how a task unfolds. These tools aim to solve the problem of how to manage multiple agents working together by providing structures that are both organized and flexible.

Overall, this direction marks a clear shift away from single-agent systems. Instead of relying on one model to do everything, researchers are building groups of agents that can specialize and collaborate. Different systems explore different ways of organizing these agents. SoT (Skeleton-of-Thought)~\cite{Ning2024SoT} speeds up communication through parallel reasoning. DMAS~\cite{Chen2023Scalable} uses a peer-to-peer setup, while ACORM~\cite{Hu2023ACORM} relies on a central controller. CAMEL~\cite{Li2023CAMEL} shows how structured prompting can guide agents to work more independently. CORY~\cite{ma2024coevolving} introduces a fine-tuning strategy based on sequential cooperative multi-agent reinforcement learning.


Despite recent advances, a fundamental limitation persists: most existing multi-agent frameworks are designed for closed or semi-centralized environments and lack explicit mechanisms to support decentralized trust.

\smallskip
\noindent\textbf{Domain-specific frameworks.}
Beyond general-purpose coordination frameworks, researchers have also developed domain-specific multi-agent systems powered by LLMs to tackle challenges in areas such as software engineering, robotics, and scientific discovery.

In software development, these systems automate tasks like coding, testing, and debugging by assigning agents to specialized roles or simulating collaborative teams through multi-persona prompting~\cite{tao2024magis,Qian2023ChatDev,Hong2023MetaGPT,Kang2023LIBRO,deng2024pentestgpt,Dong2024SelfCollaboration}. In robotics, agent teams combine language-based planning with low-level control to manage physical tasks involving manipulation, navigation, or decentralized coordination~\cite{kannan2024smart,Mandi2023RoCo,zhu2025lamarl,Zhang2023CoELA,yao2025multi}. In scientific domains, agent collectives have been used to design experiments, generate hypotheses, and engage in structured debates to improve reasoning quality and factual consistency~\cite{Bran2024ChemCrow,Ghafarollahi2024ProtAgents,Liang2023MAD}.



\subsection{Incentive Mechanisms for LLMs}

\noindent\textbf{Internal alignment incentives: reward model-based design.}
Aligning LLMs with human intent relies on transforming vague preferences into learnable incentive signals. A common approach is to use reward models that guide learning either explicitly or implicitly.

Reinforcement Learning from Human Feedback (RLHF) represents the standard pipeline, where a reward model is trained on human comparisons and then used to fine-tune the LLM through reinforcement learning algorithms such as PPO~\cite{cao2024survey, peiyuan2024agile, qu2025latent}.

To avoid the cost and instability of RL, newer methods like Direct Preference Optimization (DPO) embed the reward signal into the loss function, removing the need for a separate reward model~\cite{xu2024dpo,rafailov2023direct}. Variants such as $\delta$-DPO~\cite{wu2024dpo} and T-DPO~\cite{zeng2024trust} further refine this signal to improve alignment quality.

To provide richer feedback, recent work decomposes global rewards into smaller units. Process-level rewards guide multi-step reasoning by scoring intermediate steps~\cite{lightman2023let}, while hierarchical rewards improve long-form generation by evaluating text at sentence and paragraph levels~\cite{li2023hierarchical}. Multi-objective reward schemes also capture diverse alignment goals like helpfulness and safety using either separate models or weighted combinations~\cite{wang2024interpretable, qi2022hybrid}.

\smallskip
\noindent\textbf{External interaction incentives: game theory-based modeling.}
When LLMs interact with others in shared environments, game-theoretic models help analyze their behavior and design effective incentives. LLMs can act as rational agents in competitive or social settings. For example, they have been modeled as low-cost competitors in content markets~\cite{yao2024human}, or as strategic participants in communication games involving trust and deception~\cite{zhang2022robust,mao2025alympics,fan2024can}.

In practical applications, game structures emerge from LLM deployment. Mechanism design has been applied to ad auctions where bidders influence generation outcomes~\cite{duetting2023mechanism}, and to fine-tuning services where profit sharing and payment rules are analyzed to encourage participation~\cite{liu2025datasentinel,laufer2023finetuning}.

Self-play setups let LLMs improve by challenging each other through adversarial tasks~\cite{akata2025playing}. In peer-evaluation scenarios, game-theoretic mechanisms have been proposed to incentivize honest and high-quality feedback~\cite{duan2024gtbench}.

\section{Conclusion and Future Work}

In this work, we proposed a blockchain-integrated multi-agent coordination framework that enables trustworthy task allocation and behavior-shaping incentive alignment. Through simulation-based experiments, we demonstrated the system’s effectiveness in promoting rational agent behavior, enhancing task execution quality, and ensuring verifiable coordination through smart contracts.

While the current implementation validates our design principles under controlled settings, several limitations remain. The framework assumes structured task input and cooperative agent behavior, and it lacks real-world deployment and full economic modeling. Future extensions will focus on incorporating multi-modal task formats and deploying in real-time environments for greater realism and decentralization.

\appendices

\section{Notation Table}.
To facilitate a clear understanding of the system’s mechanisms and mathematical formulations, Table~\ref{tab:notation} summarizes all the key symbols used throughout the paper. The notation covers core concepts such as agent capabilities, task properties, reputation dynamics, utility modeling, assignment probabilities, etc. We also include relevant smart contract function names and emitted events in the table.

\begin{table}[!htbp]
\footnotesize
\centering
\caption{Notation Table for Multi-Agent Framework}
\label{tab:notation}
{%
\renewcommand{\arraystretch}{1.2}
\setlength{\tabcolsep}{5pt} 
\begin{tabularx}{\linewidth}{@{}>{\hsize=0.22\hsize}X>{\hsize=0.78\hsize}X@{}}
\toprule
\textbf{Symbol} & \textbf{Description} \\
\midrule
$A_i$ & The $i$-th agent in the system \\
$T_j$ & The $j$-th task to be assigned \\
$\mathcal{A}$ & Set of all agents in the system \\
$\mathcal{T}$ & Global taxonomy of capability tags \\
$\mathbf{r}_j$ & Binary vector for required capabilities for task $T_j$ \\
$\boldsymbol{w}_i^t$ & Continuous capability weight vector of agent $A_i$ at time $t$ \\
$\hat{\mathbf{w}}_i^t$ & Binarized capability vector derived from $w_i^t$ \\
$w_{i,k}^t$ & Weight for capability tag $k$ of agent $A_i$ at time $t$ \\
$\rho_i^t$ & Reputation score of agent $A_i$ at time $t$ \\
$\ell_i^t$ & Task workload for agent $A_i$ at time $t$ \\
$\pi_i^t(T_j)$ & Assignment probability of task $T_j$ to agent $A_i$ at time $t$ \\
$R_j$ & Reward associated with task $T_j$ \\
$U_i^t(T_j)$ & Expected utility of agent $A_i$ for task $T_j$ at time $t$ \\
$c_i^t(T_j)$ & Cost incurred by agent $A_i$ when executing task $T_j$ \\
$S_i^t$ & Task performance score received by agent $A_i$ at time $t$ \\
$q_i^t$ & Quality component of task score $S_i^t$ \\
$d_i^t$ & Delay ratio component of task score $S_i^t$ \\
$\alpha$ & Smoothing coefficient for reputation update \\
$\mu$ & Smoothing coefficient for capability weight update \\
$\beta, \gamma$ & Cost parameters for workload and skill mismatch \\
$\theta$ & Threshold for capability binarization \\
$\mathsf{CapMatch}(T_j, A_i)$ & Capability match score between task $T_j$ and agent $A_i$ \\
$\mathsf{Score}(T_j, A_i)$ & Final agent selection score for task $T_j$ and agent $A_i$ \\
\midrule
\multicolumn{2}{@{}l}{\textbf{Smart Contract Functions and Events}} \\
\midrule
$\mathsf{registerAgent}$ & Function to register an agent on-chain with metadata \\
$\mathsf{submitTask}$ & Function to publish a new task to the blockchain \\
$\mathsf{assignTask}$ & Executes task assignment policy and emits allocation event \\
$\mathsf{logAction}$ & Logs agent behavior and intermediate states \\
$\mathsf{updateReputation}$ & Updates agent's reputation based on task performance \\
$\mathsf{updateCapability}$ & Updates agent’s capability weights after task completion \\
$\mathsf{TaskAssigned}$ & Event emitted when a task is assigned to an agent \\
$\mathsf{TaskCompleted}$ & Event emitted when task execution is reported \\
\bottomrule
\end{tabularx}%
}
\end{table}



\ifCLASSOPTIONcaptionsoff
  \newpage
\fi


\bibliographystyle{IEEEtran}
\bibliography{reference}

%

%

\begin{IEEEbiography}[{\includegraphics[width=1in,height=1.25in,clip,keepaspectratio]{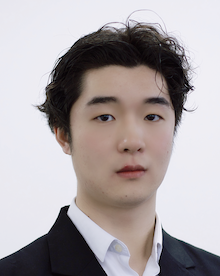}}]{Minfeng Qi} (Member, IEEE) received his master's degree in information systems from Monash University, Australia, in 2019, and his Ph.D. degree in computer science from Swinburne University of Technology, Australia, in 2023. He is currently an assistant professor at the City University of Macau. His primary research areas include blockchain security and AI security.
\end{IEEEbiography}

\begin{IEEEbiography}[{\includegraphics[width=1in,height=1.25in,clip,keepaspectratio]{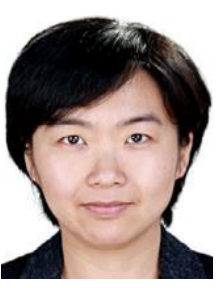}}]{Tianqing Zhu} (Senior Member, IEEE) received the B.Eng. degree in chemistry and the M.Eng. degree in automation from Wuhan University, Wuhan, China, in 2000 and 2004, respectively, and the Ph.D. degree in computer science from Deakin University, Docklands, VIC, Australia, in 2014. 

She is a Professor with the Faculty of Data Science, City University of Macau, Macau, China. Her research interests include privacy-preserving,
cybersecurity, and privacy in artificial intelligence.
\end{IEEEbiography}

\begin{IEEEbiography}[{\includegraphics[width=1in,height=1.25in,clip,keepaspectratio]{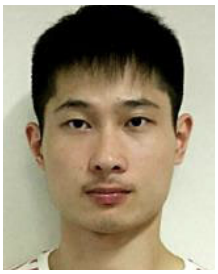}}]{Lefeng Zhang} received his B.Eng. and M.Eng. degree from Zhongnan University of Economics and Law, China, in 2016 and 2019, respectively, and a PhD degree from the University of Technology Sydney, Australia, in 2024. He is currently an assistant professor at the City University of Macau. His research interests are game theory and privacy-preserving.

\end{IEEEbiography}

\begin{IEEEbiography}[{\includegraphics[width=1in,height=1.25in,clip,keepaspectratio]{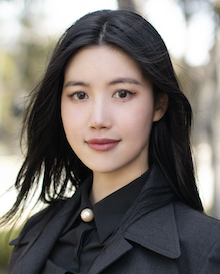}}]{Ningran Li} received the B.Sc. degree in Information Systems, the M.Sc. degree in Information Technology, and the Ph.D. degree in Computer Science from Swinburne University of Technology, Melbourne, Australia, in 2024. She is currently an Assistant Professor at the University of Adelaide, Australia. Her research interests mainly focus on cybersecurity, blockchain security, and cross-chain technologies.
\end{IEEEbiography}

\begin{IEEEbiography}[{\includegraphics[width=1in,height=1.25in,clip,keepaspectratio]{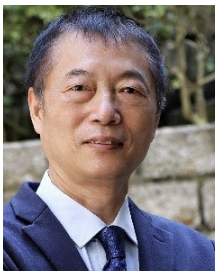}}]{Wanlei Zhou} (Fellow, IEEE) received the B.Eng. and M.Eng. degrees in computer science and engineering from Harbin Institute of Technology, Harbin, China in 1982 and 1984, respectively, the Ph.D. degree in computer science and engineering from The Australian National University, Canberra, ACT, Australia, in 1991, and the D.Sc. degree (a Higher Doctoral degree) from Deakin University, VIC, Australia, in 2002. 

He is currently the Vice Rector of the City University of Macau, Macau, China. He has published more than 400 papers in refereed international journals and refereed international conferences proceedings, including many articles in IEEE transactions and journals. His research interests include security, privacy-preserving, and distributed systems.
\end{IEEEbiography}




\end{document}